\documentclass[preprint,12pt]{aastex}
\usepackage{natbib}
\begin{document}

\title{Theoretical Models of Superbursts on Accreting Neutron Stars}
\shorttitle{Superburst Models}

\author{Randall L.\ Cooper and Ramesh Narayan}
\affil{Harvard-Smithsonian Center for Astrophysics, 60 Garden 
Street, Cambridge, MA 02138}

\email{rcooper@cfa.harvard.edu, rnarayan@cfa.harvard.edu}


\begin{abstract}

We carry out a general-relativistic global 
linear stability analysis of the amassed carbon fuel on the surface of an 
accreting neutron star to determine the conditions under 
which superbursts occur.  We reproduce the general observational 
characteristics of superbursts, including burst fluences, recurrence 
times, and the absence of superbursts on stars with accretion rates 
$\dot{M} < 0.1 \dot{M}_{\mathrm{Edd}}$, where $\dot{M}_{\mathrm{Edd}}$
denotes the Eddington limit.  
By comparing our results with observations, we are able to set
constraints on neutron star parameters such as the stellar radius and 
neutrino cooling mechanism in the core.  
Specifically, we find that accreting neutron stars with ordered crusts and 
highly efficient neutrino emission in their cores (due to direct URCA or 
pionic reactions, for example) 
produce extremely energetic ($> 10^{44}$ ergs) superbursts which are 
inconsistent with observations, in agreement with previous investigations.  
Also, because of pycnonuclear burning of carbon, they do not have 
superbursts in the range of accretion rates at which superbursts are 
actually observed unless the crust is very impure.  Stars with less 
efficient neutrino 
emission (due to modified URCA reactions, for example) produce bursts that 
agree better with observations.  
Stars with highly inefficient neutrino emission in their cores produce 
bursts that agree best with observations.  
Furthermore, we find that neutron stars with large radii ($R \sim 16$ km) 
produce very energetic superbursts that conflict with observations, even 
if the core neutrino emission mechanism is highly inefficient.
Superburst 
characteristics are quite sensitive to several other parameters as well, 
most notably the composition of the accreted gas, concentration of carbon 
in the ignition region, and degree of crystallization of the crust.  
All systems that accrete primarily hydrogen and in which superbursts 
are observed show evidence of H- and He-burning delayed mixed bursts.  
We speculate that delayed mixed bursts provide sufficient 
amounts of carbon fuel for superbursts and are thus a prerequisite 
for having superbursts.  We compare our global stability analysis to 
approximate one-zone criteria used by other authors and identify a 
particular set of approximations that give accurate results for most 
choices of parameters.  
\end{abstract} 


\section{Introduction}

Superbursts are energetic thermonuclear flashes on the surfaces of 
accreting neutron stars thought to be caused by unstable carbon 
burning \citep{WT76,TP78}.  They are similar to their 
hydrogen- and helium-burning Type I X-ray burst counterparts (which we 
refer to as ``normal'' bursts) in that they have thermal spectra, 
fast rises,  and roughly exponential decays (see Strohmayer \& Bildsten 2004 
for a review).  Superbursts distinguish 
themselves, however, by their much larger fluences and longer recurrence 
times.  Superbursts are a fairly 
new observational phenomenon:  The first superburst was discovered in 
the system 4U 1735-444 by \citet{CHKV00}.  Since then, 
eight more superbursts have been discovered in six other sources (see 
Kuulkers 2004 for a review).  All nine superbursts had observed energies 
of $\sim 10^{42}$ ergs and were detected in systems with accretion rates 
between $10 \%$ and $30 \%$ of the Eddington limit.  Recently, several 
more superburst candidates were observed in GX 17+2, which accretes at 
a rate near the 
Eddington limit \citep{intZCC04,intZetalHEAD04}.  Superburst recurrence 
times are not well constrained, though three have been observed 
within 4.7 years from the system 4U 1636-536 \citep{W01,SM02,Ketal04}.

Previous theoretical studies of superbursts \citep{BB98,CB01,SB02} have 
been quite successful 
at modeling the general properties of superbursts.  These models 
demonstrate that a large carbon-enriched layer on the surface of an 
accreting neutron star can indeed produce a
thermonuclear burst due to unstable carbon burning if the 
accretion rate $\dot{M} \gtrsim 0.1 \dot{M}_{\mathrm{Edd}}$.  The 
resulting energetics and recurrence times of the thermonuclear flashes 
are roughly consistent with observations.  However, these models employ an 
approximate criterion to estimate the column depth at which the unstable 
carbon ignition will occur.  Specifically, they define ignition to occur 
at the depth at which the 
condition $\mathrm{d} \epsilon_{\mathrm{nuc}} / 
\mathrm{d} T > \mathrm{d} \epsilon_{\mathrm{cool}} / \mathrm{d} T$ 
is satisfied, where $\epsilon_{\mathrm{nuc}}$ is the nuclear energy 
generation rate and $\epsilon_{\mathrm{cool}}$ is an approximation to 
the cooling rate.  Furthermore, these models integrate the stellar 
structure and time evolution equations only down to the superburst ignition 
region, applying  physically-motivated but not necessarily 
self-consistent boundary conditions at both ends of their computational 
domains.  However, the long characteristic accretion timescale for 
superbursts means that 
the thermal diffusion depth is quite large, often 
down to the core.  Thus, to accurately model superbursts, one must 
solve for the thermal and hydrostatic profiles of the entire crust, well
below the ignition region.  Recently, \citet{B04} constructed an improved 
model that integrates the equations all the way to the core and solves 
for the inner temperature boundary condition using a method that is more 
self-consistent than previous studies.  However, he uses the same 
approximate ignition criterion described above and assigns a fixed outer 
temperature boundary condition.   

Some authors have attempted to go beyond these simple approximations in 
modeling normal Type I X-ray bursts.  
\citet{FL87} were the first to discuss global perturbations of the 
steady-state configuration of the accreted layer.  Their analysis was 
rather crude, however, since it assumed a constant temperature 
perturbation as a function of depth and did not treat perturbations below 
the accreted layer.  
\citet[][hereafter Paper I]{NH03} developed this approach much further by 
carrying out a full global linear stability 
analysis to determine 
the ignition conditions of normal Type I X-ray bursts, thereby putting 
thermonuclear burst theory on a more rigorous footing.  However, 
their code has several limitations that, while relatively inconsequential 
for the study of normal bursts, are inadequate for the application to 
superbursts.  Explicitly, the model omits carbon burning, is not 
fully general-relativistic, and is unable to integrate the stellar 
structure and time evolution equations below the depth at which neutron drip 
occurs.  In the present investigation, we remove these limitations by 
significantly improving and expanding the burst model of Paper I.  
We then apply the new model to superbursts.

The motivation of this present study is to develop a mathematically more 
rigorous and 
self-consistent stability analysis to determine the physical conditions 
under which superbursts occur.  Our goals are to 
compare and contrast our results to preceding theoretical work, determine 
the effects of a wide range of neutron star parameters upon superburst 
characteristics, and put the theory of superburst ignition on a more 
rigorous footing 
to enable quantitative comparisons with observations.  Some of the results 
of this research were previously reported in \citet{CN04}.

We begin the paper in \S2 with a description of our numerical model, 
highlighting the additions to and improvements of the original model 
of Paper I.  In \S3, we discuss the thermal profile of the accreted layer 
and crust of the neutron star.  In particular, we illustrate the effects 
on the thermal profile of varying several physical parameters such as 
the accretion rate, accreted gas composition, accreted layer 
composition, impurity concentration, degree of ion crystallization in the 
crust, and neutrino emission mechanism in the core.  The consequences 
of these parameters upon the resulting superburst characteristics, notably 
the energetics and recurrence time, are significant, as we describe in 
\S4.  In \S5 we compare our results with observations.  We compare and 
contrast the results of our global linear stability analysis to those of 
the one-zone approximation in \S6, and we conclude with a summary in \S7.

\section{The Model}

In this section, we outline the basic theoretical model, emphasizing 
the improvements over the original model described in Paper I.  However, 
we will only briefly review the stability analysis procedure.  The reader 
is encouraged to refer to Paper I for details.

\subsection{Governing Equations}

We assume that gas accretes spherically onto a compact object of 
gravitational mass $M$ 
and areal radius $R$ at a rate $\dot{M}$, where $\dot{M}$ is the rest mass 
accreted per unit time as measured by an observer at infinity.  We consider 
all physical quantities to be functions of $\Sigma$, which we define as 
the rest mass of the accreted gas as measured from the top of the accreted 
layer divided by $4 \pi R^{2}$.  Near the stellar surface, $\Sigma$ is 
properly 
interpreted as the column density.  We use $\partial / \partial t$ and 
$\partial / \partial \Sigma$ to represent the Eulerian time and spatial 
derivatives, respectively, and $\mathrm{d} / \mathrm{d} t$ for the 
Lagrangian derivative following a parcel of gas:
\begin{equation}
\frac{\mathrm{d}}{\mathrm{d} t} = \frac{\partial}{\partial t} +
	\frac{\dot{M}}{4 \pi R^{2}} e^{-\Phi/c^{2}} \frac{\partial}
	{\partial \Sigma}.
\end{equation}
Here $\Phi$ is the metric function \citep{MTW73}, which 
reduces to the gravitational potential in the Newtonian limit.
The term $e^{-\Phi/c^{2}}$ is equal to the redshift $1 + z$ and relates 
time in the local frame to time at infinity.
We include hydrogen, helium, and carbon burning for the nuclear energy 
generation rates.  Thus, we describe the composition of the gas by the 
hydrogen mass fraction $X$, helium fraction $Y$, CNO fraction 
$Z_{\mathrm{CNO}}$ and heavy element fraction 
$Z = 1 - X - Y - Z_{\mathrm{CNO}}$, where $Z$ refers to all metals other 
than CNO.

The stellar structure and time evolution of the accreting gas are 
governed by a set of nine partial differential equations \citep{T77,B00}: 
\begin{equation}
\frac{{\partial}r}{\partial \Sigma} = - \frac{R^{2}}{m_{b} n r^{2}} 
	(1-\frac{2 G m}{r c^{2}})^{1/2},
\end{equation}
\begin{equation}
\frac{{\partial} m}{{\partial} \Sigma} = - \frac{4 \pi R^{2} \rho}
	{m_{b} n} (1-\frac{2 G m}{r c^{2}})^{1/2},
\end{equation}
\begin{equation}
\frac{{\partial} \Phi}{{\partial} \Sigma} = - \frac{GmR^{2}}{m_{b} n r^{4}}
	(1+\frac{4 \pi r^{3} P}{m c^{2}}) (1-\frac{2 G m}{r c^{2}})^{-1/2},
\end{equation}
\begin{equation}
\frac{{\partial} P}{{\partial} \Sigma} = \frac{GmR^{2}\rho}{m_{b} n r^{4}}
	(1+\frac{P}{\rho c^{2}}) (1+\frac{4 \pi r^{3} P}{m c^{2}})
	(1-\frac{2 G m}{r c^{2}})^{-1/2},
\end{equation}
\begin{equation}
e^{-2 \Phi/c^{2}} \frac{{\partial}}{{\partial}\Sigma}(\frac{F r^{2}}{R^{2}}
	e^{2 \Phi/c^{2}}) = - T 
	\frac{{\mathrm{d}}s}{{\mathrm{d}}t} - (\epsilon_{\mathrm{H}} + 
	\epsilon_{\mathrm{He}} + \epsilon_{\mathrm{C}} + 
	\epsilon_{\mathrm{N}} - \epsilon_{\mathrm{\nu}}),
\end{equation} 
\begin{equation}
e^{-\Phi/c^{2}} \frac{{\partial}}{{\partial}\Sigma}(T e^{\Phi/c^{2}}) = 
	\frac{3 R^{2} F \rho \kappa}{16 \sigma m_{b} n T^{3} r^{2}},
\end{equation}
\begin{equation}
\frac{{\mathrm{d}}X}{{\mathrm{d}}t} = - 
	\frac{\epsilon_{\mathrm{H}}}{E^{*}_{\mathrm{H}}},
\end{equation} 
\begin{equation}
\frac{{\mathrm{d}}Y}{{\mathrm{d}}t} = 
	\frac{\epsilon_{\mathrm{H}}}{E^{*}_{\mathrm{H}}} - 
	\frac{\epsilon_{\mathrm{He}}}{E^{*}_{\mathrm{He}}},
\end{equation} 
\begin{equation}
\frac{{\mathrm{d}}Z_{\mathrm{CNO}}}{{\mathrm{d}}t} = 
	\frac{\epsilon_{\mathrm{He}}}
	{E^{*}_{\mathrm{He}}} - \frac{\epsilon_{\mathrm{C}}}
	{E^{*}_{\mathrm{C}}}.
\end{equation}
Note that equation (5) is the Tolman-Oppenheimer-Volkoff equation of 
hydrostatic equilibrium.
In these equations $m$ is the interior gravitational mass, $\rho$ is the 
mass density (such that $\rho c^{2}$ is 
the energy density), $m_{b}$ is the mass of one baryon, $n$ is the baryon
number density, $r$ is the areal (Schwarzschild) radius, $P$ is the 
pressure, $F$ is the energy flux, $T$ is the temperature, $s$ is the 
entropy per unit mass, $\epsilon_{\mathrm{H}}$, $\epsilon_{\mathrm{He}}$, 
$\epsilon_{\mathrm{C}}$, and $\epsilon_{\mathrm{N}}$ are the energy 
generation rates due to 
hydrogen burning, helium burning, carbon burning, and deep crustal heating,  
$\epsilon_{\mathrm{\nu}}$ is the energy loss rate due to neutrino 
emission, $\kappa$ is the opacity, and 
$E^{*}_{\mathrm{H}}$, $E^{*}_{\mathrm{He}}$, and $E^{*}_{\mathrm{C}}$ are 
the total nuclear energies released per unit mass 
of hydrogen, helium, and carbon burned, respectively. 

Note that in the post-Newtonian stellar structure equations, the partial 
derivatives of the physical parameters are taken with respect to the 
Eulerian variable $\Sigma$, while the derivatives should be taken with 
respect to a Lagrangian variable.  However, we solve the equations in 
quasi-steady state.  Specifically, to find the equilibrium configuration, 
we set $\partial / \partial t = 0$ 
and solve the following set of ordinary differential equations
\begin{equation}
\frac{{\mathrm{d}}r}{\mathrm{d} \Sigma} = - \frac{R^{2}}{m_{b} n r^{2}} 
	(1-\frac{2 G m}{r c^{2}})^{1/2},
\end{equation}
\begin{equation}
\frac{{\mathrm{d}} m}{{\mathrm{d}} \Sigma} = - \frac{4 \pi R^{2} \rho}
	{m_{b} n} (1-\frac{2 G m}{r c^{2}})^{1/2},
\end{equation}
\begin{equation}
\frac{{\mathrm{d}} \Phi}{{\mathrm{d}} \Sigma} = - \frac{GmR^{2}}{m_{b} 
	n r^{4}}
	(1+\frac{4 \pi r^{3} P}{m c^{2}}) (1-\frac{2 G m}{r c^{2}})^{-1/2},
\end{equation}
\begin{equation}
\frac{{\mathrm{d}} P}{{\mathrm{d}} \Sigma} = \frac{GmR^{2}\rho}{m_{b} 
	n r^{4}}
	(1+\frac{P}{\rho c^{2}}) (1+\frac{4 \pi r^{3} P}{m c^{2}})
	(1-\frac{2 G m}{r c^{2}})^{-1/2},
\end{equation}
\begin{equation}
e^{-2 \Phi/c^{2}} \frac{{\mathrm{d}}}{{\mathrm{d}}\Sigma}(\frac{F r^{2}}{R^{2}}
	e^{2 \Phi/c^{2}}) = - T \frac{\dot{M}}{4 \pi R^{2}}e^{-\Phi/c^{2}}
	\frac{{\mathrm{d}}s}{{\mathrm{d}}\Sigma} - (\epsilon_{\mathrm{H}} + 
	\epsilon_{\mathrm{He}} + \epsilon_{\mathrm{C}} + 
	\epsilon_{\mathrm{N}} - \epsilon_{\mathrm{\nu}}),
\end{equation} 
\begin{equation}
e^{-\Phi/c^{2}} \frac{{\mathrm{d}}}{{\mathrm{d}}\Sigma}(T e^{\Phi/c^{2}}) = 
	\frac{3 R^{2} F \rho \kappa}{16 \sigma m_{b} n T^{3} r^{2}},
\end{equation}
\begin{equation}
\frac{{\mathrm{d}}X}{{\mathrm{d}}\Sigma} = - \frac{4 \pi R^{2}}{\dot{M}} 
	e^{\Phi/c^{2}} \frac{\epsilon_{\mathrm{H}}}{E^{*}_{\mathrm{H}}},
\end{equation} 
\begin{equation}
\frac{{\mathrm{d}}Y}{{\mathrm{d}}\Sigma} = \frac{4 \pi R^{2}}{\dot{M}} 
	e^{\Phi/c^{2}} (\frac{\epsilon_{\mathrm{H}}}{E^{*}_{\mathrm{H}}} - 
	\frac{\epsilon_{\mathrm{He}}}{E^{*}_{\mathrm{He}}}),
\end{equation} 
\begin{equation}
\frac{{\mathrm{d}}Z_{\mathrm{CNO}}}{{\mathrm{d}}\Sigma} = \frac{4 \pi R^{2}}
	{\dot{M}} e^{\Phi/c^{2}} (\frac{\epsilon_{\mathrm{He}}}
	{E^{*}_{\mathrm{He}}} - \frac{\epsilon_{\mathrm{C}}}
	{E^{*}_{\mathrm{C}}}).
\end{equation}
Clearly, since the compact object continuously accretes matter from its 
companion, the system is never precisely in equilibrium.  However, the 
accretion timescale is longer than that of most of the relevant physical 
processes, 
so the steady-state approximation is acceptable.  In this case, $\Sigma$ 
becomes a good Lagrangian variable, and so the use of $\Sigma$ as our 
independent variable is justified.

Stable and/or unstable hydrogen and helium burning produces most of 
the carbon that 
ultimately triggers a superburst.  As of this writing, each system with 
an observed superburst 
also undergoes normal Type I X-ray bursts \citep{K04,intZCC04}.  
The carbon yield resulting from both stable helium burning and unstable 
burning during normal bursts is uncertain (see \S4).  
To account for this uncertainty, we introduce a free parameter 
$C_{\mathrm{f}}$, the fraction of hydrogen and helium that ultimately 
burns to carbon.  Equation (19) thus becomes 
\begin{equation}
\frac{{\mathrm{d}}Z_{\mathrm{CNO}}}{{\mathrm{d}}\Sigma} = \frac{4 \pi R^{2}}
	{\dot{M}} e^{\Phi/c^{2}} (C_{\mathrm{f}} 
	\frac{\epsilon_{\mathrm{He}}}
	{E^{*}_{\mathrm{He}}} - \frac{\epsilon_{\mathrm{C}}}
	{E^{*}_{\mathrm{C}}}),
\end{equation}
so at the base of the accreted layer, 
$Z_{\mathrm{CNO}} \approx C_{\mathrm{f}}$.  Clearly this is an 
approximation.  However, it enables us to model the composition of the 
accreted layer as well as possible.  Specifically, we can model 
the composition of both the hydrogen- and helium-rich upper region of the 
outer crust and the carbon-rich lower region to high accuracy.  
Additionally, the layer in which the hydrogen and helium burn 
to carbon is most likely very narrow in the column density $\Sigma$, so this 
transition region is inconsequential to the thermal and hydrostatic 
profiles of the outer crust.

\subsection{Boundary Conditions}

The solution to the set of nine coupled partial differential equations 
(11-19) requires nine separate boundary conditions.  Eight are applied at 
the photosphere (where the optical depth $\tau_{\mathrm{out}}$ = 2/3) and 
one is applied at the crust-core interface.

The outer boundary conditions for equations (11-13) are $r = R$, $m = M$, 
and $\Phi = (c^{2}/2) \ln(1-2 G M / R c^{2})$.  
The outer boundary conditions for the hydrogen, helium, and CNO mass 
fraction evolution equations (17-19) are given by the composition of the 
accreting 
gas, such that $X = X_{\mathrm{out}}$, $Y = Y_{\mathrm{out}}$, and
$Z_{\mathrm{CNO}} = Z_{\mathrm{CNO, out}}$, respectively.  The value for 
$\Sigma$ at the photosphere, $\Sigma_{\mathrm{out}}$, is 
obtained approximately by taking the opacity at the photosphere to be 
given by electron scattering.  Thus, $\Sigma_{\mathrm{out}} = 
\tau_{\mathrm{out}}/(0.2 (1 + X_{\mathrm{out}}))$ $\mathrm{g\,cm}^{-2}$.  
The outer boundary condition for equation (14) is then given by 
\begin{equation}
P_{\mathrm{out}} = \Sigma_{\mathrm{out}} \frac{G M}{R^{2}} (1 - 
	\frac{2 G M}{R c^{2}})^{-1/2}.
\end{equation}

The method we use to determine the outer boundary condition for equation 
(15), the outward flux at the stellar surface $F_{\mathrm{out}}$, is 
explained in Paper I.  To summarize, we assume a given value 
of $F_{\mathrm{out}}$ which, when added to the gravitational energy 
flux due to accretion,  also defines a surface temperature 
$T_{\mathrm{out}}$.  We then integrate the differential equations and 
compare the resulting temperature at the bottom, the crust-core interface, 
to the required temperature inner boundary condition.  We adjust 
$F_{\mathrm{out}}$ and repeat the integration until 
the temperature boundary condition at the bottom is satisfied to high 
accuracy.  What makes our new method 
superior to that of Paper I is that we now integrate all the 
way to the stellar core ($\rho_{0} \approx 2 \times 10^{14}$ 
$\mathrm{g\,cm^{-3}}$, where $\rho_{0}$ denotes the rest mass density).  
Previously, we integrated only a couple of diffusion depths into the star, 
but we were unable to integrate past the neutron drip point 
($\rho_{0} \approx 4 \times 10^{11}$ $\mathrm{g\,cm^{-3}}$).  The long 
recurrence times of superbursts make the old method inadequate.  
For many calculations, the thermal diffusion depth is deeper than the 
crust-core interface, so integration to the core is necessary.

For a given calculation, we employ one of two methods to determine the 
temperature inner boundary condition for equation (16).   
In our first method, we assume that we know the 
rate of neutrino emission from the core; we use two prescriptions for this, 
either modified 
URCA reactions \citep{FM79,YL95} or pionic reactions \citep{MBCDM77}.  
Integrating to the crust-core 
interface gives values for the proper temperature $T$, areal radius $r$, 
interior gravitational mass $m$, and
energy flux $F$.  The energy flux, which is directed inward at 
the interface ($F < 0$), must be balanced by the neutrino 
cooling of the core.  Using the neutrino luminosities 
$L_{\nu}(m,T)$ from \citet{ST83}, we determine the core temperature 
$T_{\mathrm{core}}$ via the equation
\begin{equation}
L_{\nu}(m,T_{\mathrm{core}}) = - 4 \pi r^{2} F.
\end{equation}
We then compare $T_{\mathrm{core}}$ to $T$ and iterate until they match.
Note that we do not modify the formulae to account for the volume of the 
core.  Since we do not know the equation of state of the core, 
we cannot solve for the core's proper volume.  However, since the 
formulae are estimates, and since the luminosity is a very strong function 
of the temperature ($L_{\nu}^{\mathrm{mURCA}} \propto T^{8}$ and 
$L_{\nu}^{\mathrm{\pi}} \propto T^{6}$), this approximation is 
innocuous. 
In our second method, we simply assign a value for the core temperature 
$T_{\mathrm{core}}$ and iterate until $T = T_{\mathrm{core}}$ at the 
crust-core interface.

\subsection{Auxiliary Conditions}
\subsubsection{Equation of State}

We assume photons, electrons, ions, and free neutrons supply the pressure.  
To calculate the photon and electron contributions to the pressure, we 
follow the same procedure as in Paper I, while for the ions and free 
neutrons, 
we follow the prescriptions of \citet{B00}.  The phase of the ions, 
whether solid or liquid,  
is determined by the dimensionless coupling parameter 
\begin{equation}
\Gamma = \frac{(Z e)^{2}}{k_{\mathrm{B}} T} (\frac{4 \pi}{3} n_{I})^{1/3}, 
\end{equation}
where $n_{I}$ is the ion number density.  In the outermost layers of the 
crust, where $\Gamma < 1$, we use the free energy fits of \citet{CP98} 
for the ion equation of state.  For $\Gamma > 1$ we use the analytical 
fits of 
\citet{FH93} for the Helmholtz free energy to calculate the ion pressure.  
We assume the ions are in the liquid phase when $1 < \Gamma < 173$ and 
the solid phase when $\Gamma > 173$.  The analytical fits 
are valid for a one-component plasma.  However, in the accreted layer more 
than one species is usually present at a given density.  In this case 
we approximate the multi-component mixture as a one-component plasma by 
making the substitutions $Z \rightarrow \langle Z \rangle$ and 
$n_{I} \rightarrow n / \mu_{nuc}$, where $\mu_{nuc}$ is the mean molecular 
weight per nucleus.
Below the accreted layer, we assume only one species is present at a given 
density.
We calculate the free neutron pressure from the compressible liquid-drop 
nuclear model of \citet{MB77}.  For $\rho_{0} > 1.3 \times 10^{13}$ 
$\mathrm{g\,cm}^{-3}$, we use the equation of state from \citet{NV73}.

The photon entropy formula is the same as that in 
Paper I.  We calculate the Coulomb portion of the ion entropy from the free 
energy fits of \citet{C93} and the ideal portion from the fits of \citet{FH93}.
The treatment of the electron and free neutron entropies is more difficult.  
The compressional heating terms in equation (15) are unlikely to 
affect the thermal profile calculation significantly \citep{BB98} and 
therefore are often omitted in other investigations.  However, we 
attempt to make a 
reasonable approximation to the compressional heating terms relevant 
to our work.  We consider three entropy ``regimes'': 
ideal, nonrelativistic degenerate, and extremely relativistic degenerate.  
Define $\Theta = m_{\mathrm{p}} k_{\mathrm{B}} T / (\hbar^{2} n^{2/3})$ 
and $x = (\hbar / m_{\mathrm{p}} c)(3 \pi^{2} n)^{1/3}$, the 
relativity parameter 
\citep{ST83}, where $m_{\mathrm{p}}$ is the particle mass, 
$k_{\mathrm{B}}$ is Boltzmann's constant, and $n$ is the particle 
number density.  Then the expressions for the entropy of a species in 
each of the three regimes are \citep{LL69}
\begin{equation}
\frac{s_{\mathrm{ideal}}}{k_{\mathrm{B}}} = \frac{3}{2} \ln(\Theta) +
\ln(\frac{2}{(2 \pi)^{3/2}}) + \frac{5}{2},
\end{equation}
\begin{equation}
\frac{s_{\mathrm{NRD}}}{k_{\mathrm{B}}} = (\frac{\pi}{3})^{2/3} \Theta = 
\frac{\pi^{2} k_{\mathrm{B}} T}{m_{\mathrm{p}} c^{2} x^{2}},
\end{equation}
\begin{equation}
\frac{s_{\mathrm{ERD}}}{k_{\mathrm{B}}} = 
\frac{\pi^{2} k_{\mathrm{B}} T}{m_{\mathrm{p}} c^{2} x}.
\end{equation}
Note that $\Theta \gg 1$ for an ideal gas and $\Theta \ll 1$ for a 
degenerate gas.  Similarly, $x \gg 1$ for a relativistic gas and $x \ll 1$ 
for a nonrelativistic gas.  The question is, then, how does one decide where 
to make the transitions between the three regimes?  It is evident from 
equation (15) that only the entropy derivatives, not the entropy 
values, are important.  Therefore, we choose the transition values 
of $\Theta$ and $x$ in such as way as to make the partial derivatives of the 
entropies with respect to the variable $\Theta$ or $x$ continuous 
functions of that 
variable.  First, we decide whether the species is ideal or degenerate.  
Thus, for a 
quantity $q$, we first find the transition value $\Theta_{\mathrm{trans}}$ 
such that $\partial s_{\mathrm{ideal}} / \partial q = 
\partial s_{\mathrm{NRD}} / \partial q$ at $\Theta = 
\Theta_{\mathrm{trans}}$. 
For all quantities $q$, $\Theta_{\mathrm{trans}} = 1.455$.  If the 
species is degenerate, we then decide whether it is nonrelativistic or 
extremely relativistic.  We then repeat the same procedure with 
$\Theta \rightarrow x$.  $x_{\mathrm{trans}} 
= 0.5$, $1$, or $2$, depending on the specific quantity $q$.

\subsubsection{Nuclear Energy Generation Rates}

The hydrogen and helium nuclear energy generation rates are identical to 
those in Paper I with the exceptions that the $^{13}$N$(p,\gamma)^{14}$O 
and $^{14}$N$(p,\gamma)^{15}$O rates for $\epsilon_{\mathrm{H}}$ are updated 
to those of \citet{CF88}.

Carbon burns through several different energetically possible reaction 
channels \citep{C83}.  For simplicity, we assume that 
the dominant reaction is $^{12}$C($^{12}$C,$\gamma$)$^{24}$Mg, so 
$E^{*}_{\mathrm{C}} = 5.6 \times 10^{17} \mathrm{erg\,g}^{-1}$ 
\citep{CB01}.  For 
$\epsilon_{\mathrm{C}}$, we use both the reaction rates and electron 
screening enhancement factors of \citet{K00}.  We interpolate the logarithms 
of the fluid and solid phase reaction rates in the same way that \citet{K00} 
interpolates the thermonuclear and pycnonuclear enhancement factors.  
This makes the energy generation rate a smooth function of both density 
and temperature.
See Figure 1 for a plot of $\epsilon_{\mathrm{C}}$ as a function of 
$\rho_{0}$.  Note that we use the approximate criterion of \citet{K00} 
for the solidification of an electron-screened binary ionic mixture.  
Previous authors \citep{BB98,CB01,SB02,B04} have used the 
reaction rate of \citet{CF88} with the 
enhancement factor of \citet{OIVH93}.  This energy 
generation rate is valid only in the liquid phase, where the burning is 
considered strictly thermonuclear.  In contrast, our energy generation rate 
is valid in both the thermonuclear regime, where the crust is fluid, 
and the pycnonuclear regime, where the crust is solid.
The two energy generation rates agree to sufficient precision in the 
thermonuclear regime.  
In the pycnonuclear regime, the nuclear energy generation rate is 
essentially temperature-independent.  Including this regime in the 
model can sometimes be important, especially 
in neutron stars with efficient core cooling mechanisms or 
heavy-element oceans.

\subsection{Deep Crustal Heating and Neutrino Emission}

For $6 \times 10^{11}$ $\mathrm{g\,cm}^{-3} \lesssim \rho_{0} \lesssim 3 
\times 10^{13}$ $\mathrm{g\,cm}^{-3}$, electron captures, neutron 
emissions, and pycnonuclear reactions 
release $\approx 1$ MeV per baryon over a timescale inversely proportional 
to the accretion rate \citep{HZ90v229,HZ90v227,BBR98}.  For 
$\epsilon_{\mathrm{N}}$, 
the energy generation rate due to this deep crustal heating, we use the 
formula of \citet{B00}.

We include pair, photo-, plasma, and bremsstrahlung neutrino emission 
processes for the neutrino energy loss rate $\epsilon_{\nu}$.  For the 
pair, photo-, and plasma processes we use the analytical fits of 
\citet{IHNK96}, and for the bremsstrahlung emission rate we use 
the work of \citet{HKY96} when the ions are in the liquid phase and 
\citet{YK96} when the ions are in the solid phase.

\section{Thermal Structure of the Accreted Layer and Crust}

Superburst characteristics depend sensitively upon the thermal structure 
of the neutron star crust.  In this section, we identify some of the 
key parameters that affect the thermal profile.  

\subsection{Accretion Rate}

In its journey from the binary companion to the neutron star core, a 
parcel of matter releases a tremendous amount of energy, and the rate of 
this energy release is directly proportional to the rate at which matter 
falls onto the stellar surface.  Thus, it is not surprising that the thermal 
profile of an accreting neutron star is very sensitive to the accretion 
rate.  To varying degrees of importance, the accretion rate affects the 
thermal profile in four different ways.   

Most of the energy released by an infalling parcel of accreted material 
is from the gravitational energy released when the matter impacts the 
stellar surface.  Most of this energy is radiated outward.  Nevertheless, 
it determines the temperature at the stellar surface and thereby sets a 
boundary condition.  The calculation is rather insensitive to this 
temperature, however, for the thermal profile near the surface 
approaches a radiative-zero solution.  
As accretion continues, nuclear fuel accumulates and eventually burns 
either stably or unstably.  The time-averaged rate of nuclear energy 
generation is proportional to the 
accretion rate.  Although the nuclear energy per gram of accreted 
material released via fusion is roughly forty times less than that released  
from gravitational energy, the nuclear energy is generated well below 
the stellar surface.  Thus the nuclear energy generation can have a 
significant effect upon the thermal profile of the crust, in particular 
the superburst ignition region (see also \S3.2).  
Additionally, continuous accretion causes both compressional heating 
throughout the crust and deep crustal heating via non-equilibrium 
reactions.  The compressional heating is rather small compared to other 
sources, and it is therefore often neglected in other studies.  The deep 
crustal heating is roughly five times less than that from hydrogen and 
helium burning, but it can have a non-negligible effect upon the thermal 
profile of the crust, especially if the conductive opacity of the inner 
crust is large.  

We plot the temperature and flux profiles for two neutron 
stars accreting at different rates in Figure 2.  The energy flux 
is normalized by the maximum nuclear burning energy flux available in the 
accreting gas:
\begin{equation}
F_{\mathrm{nuc}} = \frac{\dot{M}}{4 \pi R^{2}} 
(1 - \frac{2 G M}{R c^{2}})^{-1/2} [X_{\mathrm{out}} E^{*}_{\mathrm{H}}
+ (X_{\mathrm{out}} + Y_{\mathrm{out}}) E^{*}_{\mathrm{He}} + 
(X_{\mathrm{out}} + Y_{\mathrm{out}} + Z_{\mathrm{CNO,out}}) 
E^{*}_{\mathrm{C}}].
\end{equation}
The parameter $l_{\mathrm{acc}}= \dot{M} / \dot{M}_{\mathrm{Edd}}$ is the 
accretion rate normalized to the Eddington limit, where 
$\dot{M}_{\mathrm{Edd}} = 4 \pi G M (1+z)/c z \kappa_{\mathrm{es}}$, 
with $\kappa_{\mathrm{es}} = 0.4$ $\mathrm{cm}^{2}\,\mathrm{g}^{-1}$.
The temperature outer boundary condition, which is shown at the left end 
of the left panel, is determined by the rate of gravitational energy 
liberated at the surface.  Changes in the slope of the temperature profile 
are associated with localized energy sources, and they are reflected by 
rapid changes in the flux.  Thus, hydrogen and helium burning causes the 
peak in the thermal profile and the large change of flux at a column depth 
$\Sigma \approx 10^{8}$ $\mathrm{g\,cm}^{-2}$.  A small amount of carbon 
burning occurs at $\Sigma \sim 10^{12}$-$10^{13}$ 
$\mathrm{g\,cm}^{-2}$.  Deep crustal heating occurs for 
column depths 
$10^{15}$ $\mathrm{g\,cm}^{-2} \lesssim \Sigma \lesssim 10^{17}$ 
$\mathrm{g\,cm}^{-2}$.  The flux profile shows that most of 
the energy generated by deep crustal heating is directed inward, in 
agreement with \citet{B00}.

\subsection{Energy Generated from Hydrogen and Helium Burning}

The burning of hydrogen and helium near the surface of an accreting 
neutron star releases a substantial amount of energy within the star.  
Consequently, the thermal profile of the outer crust, including the 
superburst ignition region, is rather sensitive to the magnitude and 
physical location of this energy generation.  To account for this, 
previous authors set the temperature at a given 
column depth to coincide with estimates from investigations of hydrogen 
and helium ignition.  However, the thermal profile in this region is a 
sensitive function of many variables, including the mass accretion rate 
(e.g., compare the two models in Fig.\ 2), 
stellar radius,  and composition of the accreted gas.  Since we include 
both hydrogen 
and helium energy generation rates in our energy conservation equation, 
we make no assumptions regarding the temperature at a given depth in the 
accreted layer.  Thus, we are able to determine the thermal profile of 
the outer crust self-consistently.  Not only does this improve the 
accuracy of our calculation, but it also gives us the freedom to vary 
physical parameters such as the gas composition and stellar radius 
self-consistently.

As noted earlier, all of the systems in which astronomers have observed 
superbursts exhibit normal Type I X-ray bursts as well.  To do a rigorous 
calculation of the thermal profile of the outer crust, one would need 
to conduct a fully time-dependent calculation of many successive normal 
bursts, which is beyond the scope of this study.  Our calculation is 
quasistatic, so the composition and thermal profile of the crust in the 
the normal burst ignition region is essentially computed via stable 
(though rapid) hydrogen and helium burning.  Since the 
timescale over which normal bursts occur (hours to days) is much shorter 
than the timescale over which superbursts occur (years to possibly 
decades), any 
effects that hydrogen and helium burning have on the thermal 
profile of the superburst ignition region will be due to the time-averaged 
hydrogen and helium nuclear energy generation rate.  The time-averaged 
energy generation rate is the same regardless of the manner in which the 
fuel is burned.  Therefore, our method should be sufficiently accurate 
for our purposes.

To demonstrate the importance of hydrogen and helium burning on the thermal 
profile of the outer crust, we plot in Figure 3 the temperature as a 
function of column density for two systems with different accreted gas 
compositions.  Hydrogen burning releases much more energy per gram of 
accreted fuel than helium burning.  Therefore, the maximum temperature 
achieved in the outer crust is in general proportional to the mass 
fraction of hydrogen of the accreted gas.  This can have a significant 
effect on superburst characteristics (see \S4.1).  The inner crust 
($\Sigma \gtrsim 10^{15}$ $\mathrm{g\,cm}^{-2}$), however, 
is rather insulated from the hydrogen and helium burning region, so the 
thermal profile of the inner crust is rather insensitive to the hydrogen 
and helium burning near the surface.  

\subsection{Ash Composition from Hydrogen and Helium Burning}

The heavy element composition of the accreted layer where superbursts are 
triggered is 
uncertain.  Especially for mixed hydrogen/helium accretors, 
the primary cause of this uncertainty is due to the rp-process 
\citep{WW81}, whose ashes are probably 
a mix of elements beyond the iron peak \citep{Setal01}.  Type I X-ray 
burst models of \citet{Wetal04} produce nuclei with an average 
atomic weight $\langle A \rangle \approx 64$.  Therefore, 
we choose two representative metals as our heavy elements: (i) 
$^{56}_{26}$Fe,  
and (ii) following \citet{CB01}, $^{104}_{~44}$Ru. These choices 
likely bracket the true average atomic weight and electric charge 
of the ashes.  We assume that 
the nuclear composition below the accreted layer is that of 
\citet{HZ90v229,HZ90v227} for 
both cases.  \citet{HZ03} later studied the evolution of heavy ($\langle 
A \rangle\approx 100$) rp-process ashes.  However, \citet{SBC03} showed 
that the high temperatures reached during a superburst might induce 
photodisintegration reactions in the heavy ashes, converting them to iron
group elements.  Thus, we presume that the outer crust below the accreted 
layer consists chiefly of iron group elements, with possible impurities.

We find that, contrary to the study of \citet{CB01}, the composition of 
the ashes from hydrogen and helium 
has a negligible effect on the thermal profile of the crust.  This is 
consistent with the results of \citet{B04}.  See Figure 4 for 
a plot of the thermal profile for neutron star crusts with different 
ash compositions.  

\subsection{Core Temperature}

The thermal profile of both the inner crust and the outer crust below the 
hydrogen/helium burning region is very sensitive to the temperature 
at the crust-core interface.  See Figure 5 for a plot of the thermal 
profiles of neutron stars with different core temperatures.  In 
particular, note the slopes of the profiles beyond the hydrogen/helium 
burning region, at $\Sigma \approx 10^{8}$ $\mathrm{g\,cm}^{-2}$.  
Except for neutron stars with very hot cores 
($T \gtrsim 4 \times 10^{8}$ K), the temperature gradient 
$\partial T / \partial \Sigma$ in this region is negative (i.e.\ 
$\partial T / \partial r$ is positive), 
which implies that the net energy flux is negative.  This is clearly 
illustrated in the energy flux profiles in Figure 5.  Previous studies of 
thermonuclear burst ignition have often used  the result of \citet{B00} 
for the inner flux boundary 
condition.  He found that approximately $10 \%$ of the 
energy generated in the crust through 
deep crustal heating flows outward.  In Figure 5 this corresponds to 
$F/F_{\mathrm{nuc}} \approx 0.02$.  For comparison, the 
energy flux in Figure 5 just below the column depth at which the 
superburst is 
triggered ranges from $F/F_{\mathrm{nuc}} \approx +0.05$ at 
$\Sigma \approx 10^{11.7}$ $\mathrm{g\,cm}^{-2}$ for the hottest 
thermal profile to $F/F_{\mathrm{nuc}} \approx -0.06$ at 
$\Sigma \approx 10^{14.0}$ $\mathrm{g\,cm}^{-2}$ for the coldest.  
Brown's results, which are accurate for the particular 
physical scenario he studied and which we are able to reproduce, 
are clearly not applicable for all 
scenarios.  As we have shown, the thermal profile of an accreting 
neutron star is sensitive to many parameters, including the accretion 
rate, composition of the accreted gas, and core temperature.  Therefore, 
one must use caution when implementing flux or temperature 
boundary conditions.  In our opinion, it is preferable to do 
self-consistent calculations, as in this work.  

In recent work, \citet{B04} has used an approximate ``outer'' boundary 
condition on the temperature: $T = 2.5 \times 10^{8}$ K at 
$\Sigma = 10^{9}$ $\mathrm{g\,cm}^{-2}$.  A quick look at Figures 2, 3, and 
5 here shows that the temperature varies considerably at this depth, 
depending on various parameters.  Also, $T$ is typically greater 
than $2.5 \times 10^{8}$ K at this depth.

\subsection{Ion Impurities in the Crust}

Electron-impurity scattering \citep{IK93} and electron-phonon scattering 
\citep{BY95} determine the thermal conductivity of the solidified inner 
crust of an accreting neutron star.
Ion impurities embedded in a crystalline lattice cause deviations in an 
otherwise periodic potential.  Electron scattering off of these deviations 
reduces the thermal conductivity of the lattice and consequently raises 
the temperature of the inner crust.  Unlike electron-phonon 
scattering, which is due to thermal oscillations of ions and is therefore 
temperature-dependent, electron-impurity scattering depends primarily on the 
lattice composition and thus the corresponding collision frequency is 
essentially temperature-independent.  Therefore, impurities can have 
a significant effect on the thermal conductivity of the crust, 
especially at low temperatures \citep{IK93,AM76}.   

The contribution of impurities to the thermal conductivity of the crust 
is parametrized by the impurity parameter \citep{IK93,B00}
\begin{equation}
Q \equiv \frac{1}{n_{I}} \sum_{j} n_{j}(Z_{j} - \langle Z \rangle)^{2},
\end{equation}
where $n_{j}$ and $Z_{j}$ are the number density and charge of the 
$j$th ion species, $n_{I} = \sum_{j} n_{j}$ is the total ion number 
density, and $\langle Z \rangle = n_{I}^{-1} \sum_{j} n_{j} Z_{j}$ is 
the mean charge.  Sources of impurities in the crust include electron 
captures and pycnonuclear reactions \citep{HZ90v229,HZ90v227}, but the 
main source of impurities is most likely the mixture of superburst and/or 
rp-process ashes at the top of the substrate.  Early studies of the 
composition of rp-process ashes implied that impurity scattering should 
be significant, with $Q \sim 100$ \citep{Setal99}.  Subsequent calculations 
by \citet{Wetal04} and \citet{KHKF04} showed that rp-process ashes consist 
chiefly of iron-peak nuclei, resulting in a lower $Q$.  \citet{SBC03} 
found that $Q \approx 5.2$ after a superburst is triggered.  In our 
model, we calculate $Q$ self-consistently within the accreted layer, and we 
adopt $Q = 5.2$ as our fiducial impurity parameter value in 
the substrate below the accreted layer.  Figure 6 
illustrates the effects of electron-impurity scattering on the thermal 
profile of an accreting neutron star.  

\subsection{Ion Crystallization in the Crust}

When calculating the conductive opacity in the inner crust of the neutron 
star, we usually assume that the ions form an ordered crystal lattice 
when the dimensionless coupling parameter $\Gamma > 173$ (eq.\ 23).  In this 
case, the conductive opacity is usually dominated by electron-phonon 
scattering 
due to oscillations of the ions in the lattice \citep{PBHY99}.  
However, previous 
studies of the nuclear structure of the inner crust suggest that some 
fraction of the inner crust may in fact be disordered \citep{MH02,MB04}.  
Following \citet{B04}, we investigate the thermal profile of a completely 
disordered neutron star crust.  We use the thermal conductivity expression 
of \citet{IK93} and set the structure factor $\langle S \rangle$ to unity.  
Additionally, we set the squared impurity charge 
$\langle ( \Delta Z )^{2} \rangle = \langle Z \rangle^{2}$.  This 
essentially sets a lower limit on the thermal conductivity of the crust 
(E.~Brown, private communication).  
The thermal profiles of the crusts for neutron stars with cores that emit 
neutrinos via either modified URCA reactions or pionic reactions are shown 
in Figure 7.  In contrast to neutron stars with crystalline 
crusts (Fig.\ 5),  the thermal profiles now are quite insensitive to the 
nature of the core neutrino cooling mechanism.  

\section{Results}

\subsection{Composition of the Accreted Gas}

For systems that have exhibited a superburst and for which the composition 
of the accreted gas can be reasonably estimated, all but one  
accrete a mixture of hydrogen and helium, with hydrogen being the most 
abundant species by mass \citep{K04}.  The system 4U 1820-30 
\citep{S00,SB02} 
is the exception.  Several studies \citep{FE89,PRP02,C03} imply that 
the compact star in this system accretes a helium-rich mixture with a 
small hydrogen 
mass fraction $X \sim 0.1$.   Observationally, the superburst from 4U 
1820-30 is distinct, with a larger fluence, luminosity, 
and peak temperature than every other superburst observed thus far 
\citep{Ketal02,K04}.  
Therefore, to determine the effects of 
accreted gas composition on superbursts, we choose two different 
elemental abundances: (i) ``mixed hydrogen/helium'', for which the mass 
fractions of 
the accreted gas are $X = 0.7$, $Y = 0.28$, and $Z_{\mathrm{CNO}} = 0.016$, 
and (ii) ``helium'', for which $X = 0.1$, $Y = 0.88$, and 
$Z_{\mathrm{CNO}} = 0.016$.  

Figure 8 shows the superburst energies and recurrence times as a 
function of accretion rate, as calculated by our model.  All other 
parameters being equal, helium accretors require a larger column 
density of accreted gas before a superburst is triggered.  Therefore, 
their superbursts are more energetic and have longer recurrence times, 
in agreement with the observations of 4U 1820-30.  
Helium burning releases less energy per gram than hydrogen burning by 
approximately one order of magnitude.  
Consequently, for a given column density, the temperature at the base 
of the layer is lower (see Fig.\ 3), so a larger column of fuel must 
accumulate before 
a superburst can occur.  This disparity is greater at high accretion rates, 
for which a lower column density is required for an instability.  Less 
matter exists between the hydrogen/helium burning region and the base of 
the layer, so the base is less insulated from the burning region and 
therefore more sensitive to the energy generated there.

As we have shown, the composition of the accreted gas affects superburst 
characteristics by its effect upon the thermal profile of the outer crust 
where superbursts are triggered.  However, the composition may have an 
even greater influence on superburst characteristics through the carbon 
yield resulting from both the stable and unstable burning of the gas.  
Unfortunately, we cannot investigate this aspect of the problem with our 
model since we do not solve for the carbon yield self-consistently, 
but set it through the parameter $C_{\mathrm{f}}$.  We assume the 
fiducial value $C_{\mathrm{f}} = 0.3$ for all calculations unless 
notified otherwise.

\subsection{Composition of the Ashes in the Accreted Layer}

The composition of the accreted layer where superbursts are triggered
is uncertain.  In particular, the mass fraction of carbon produced via
stable and unstable hydrogen and helium burning is unknown, but it most
likely has to be $\gtrsim10\%$ for mixed hydrogen/helium accretors
\citep{CB01} and $\gtrsim30\%$ for helium accretors \citep{SB02}.
These authors find that, for $\dot{M} \lesssim 0.3
\dot{M}_{\mathrm{Edd}}$, a smaller mass fraction of carbon than these
limits will not produce a superburst.  Especially for mixed
hydrogen/helium accretors, the primary cause of the uncertainty in the
carbon fraction is due to the rp-process \citep{WW81}.  Several
research groups have studied the final products of the rp-process from
both stable and unstable burning \citep{SBCW99,SBCO03,KHKF04,Wetal04,
Fetal04}.  They
find that the carbon mass fraction of the ashes is notably below
$10\%$ for the range of accretion rates at which superbursts have been
observed, which is problematic since such a low fraction is
insufficient to trigger a superburst.  The exception to this statement
may be the system GX 17+2, in which extremely energetic Type I X-ray
bursts were observed at accretion rates in the neighborhood of the
Eddington limit \citep{KHvdKLM02,intZCC04}.  Despite these results, we
cannot rule out the possibility that some other process (e.g., 
delayed mixed bursts, see \S5) may still be able to produce enough carbon.  
Due to this uncertainty, we take as a free
parameter $C_{\mathrm{f}}$, the fraction of hydrogen and helium that
ultimately burns to carbon.

Figure 9 shows superburst energies and recurrence times as a function of 
accretion rate for three choices of $C_{\mathrm{f}}$.  For a 
given accretion rate, the superburst recurrence time will usually be 
shorter for 
a larger mass fraction of carbon at the base of the accreted layer because 
an instability will occur sooner if more carbon is present. To 
first order, the burst energy $E_{\mathrm{burst}} \propto 
\Sigma_{\mathrm{layer}} C_{\mathrm{f}}$, where $\Sigma_{\mathrm{layer}}$ 
is the column depth of the accreted layer when a superburst is triggered.
However, $\Sigma_{\mathrm{layer}}$ itself is roughly proportional to the 
recurrence time, so it is difficult to derive a general relationship 
between $C_{\mathrm{f}}$ and $E_{\mathrm{burst}}$ (compare the burst 
energies for $C_{\mathrm{f}} = 0.3$ and $C_{\mathrm{f}} = 0.5$ in 
Figure 9).  In general, the value of $C_{\mathrm{f}}$ does not 
substantially affect superburst energetics and recurrence times for a 
given accretion rate.  However, the value of $C_{\mathrm{f}}$ does 
significantly affect the lower limit of the range of accretion rates 
at which superbursts occur.  We find that the lower limit is roughly 
inversely proportional to $C_{\mathrm{f}}$, in agreement with 
\citet{CB01}.

As noted in \S3.3, the composition of the heavy elements in the accreted 
layer has a negligible affect on the thermal profile of the superburst 
ignition 
region.  Consequently, for accretion rates at which superbursts are 
triggered, superburst energies and recurrence times are quite 
insensitive to the nuclear composition of the heavy elements in the 
accreted layer.  This agrees well with the results of \citet{B04}.  At lower 
accretion rates, however, the greater charge of the $^{104}_{~44}$Ru ions 
may cause the base of the accreted layer to solidify before a thermonuclear 
instability can occur.  In this case, the carbon fuel burns stably via 
pycnonuclear reactions (see \S2.3.2), so a superburst does not occur.  
Therefore, though 
the composition of the heavy elements in the ocean is unimportant with 
regard to  superburst characteristics, it is important with regard to the 
presence or absence of superbursts at a given accretion rate.  This
dependence has not been noted in previous studies.  See Figure 10 
for a plot of superburst energies and recurrence times for neutron star 
crusts with different heavy element compositions.

\subsection{Neutrino Cooling Mechanism in the Core}

The composition of dense matter in the inner cores of neutron stars 
is essentially unknown.  However, this composition significantly affects 
the neutrino emission there.  Thus, knowledge of which neutrino 
processes occur in the core can help constrain the types of 
matter that exist at such high densities.  

When we solve for the core temperature by balancing the flux flowing inward 
with the neutrino cooling, we find 
$T_{\mathrm{core}} \approx 3 \times 10^{8}$ K for nonsuperfluid 
cores that emit neutrinos via modified URCA reactions and 
$T_{\mathrm{core}} \approx 2 \times 10^{7}$ K for cores 
that emit neutrinos via pionic reactions.  In general, $T_{\mathrm{core}} 
\propto \dot{M}^{1/8}$ for modified URCA cooling and $T_{\mathrm{core}} 
\propto \dot{M}^{1/6}$ for pionic cooling.
The thermal profile of much of the crust, and in particular the region 
in which superbursts are triggered, is quite sensitive to the core 
temperature (see Fig.\ 5).  Thus, compared to stars with inefficient 
cooling mechanisms 
in their cores, neutron stars with efficient cooling mechanisms in their 
cores need to 
accrete more fuel in order for the carbon at the base of the accreted layer 
to ignite.  Consequently, these superbursts will be more energetic and will 
have longer recurrence times, as shown in Figure 11.  

If neutrons and protons exist in the stellar core, they probably
become superfluid at high densities \citep{BPP69}.  Baryon
superfluidity drastically suppresses neutrino emission from modified
and direct URCA processes and therefore raises the temperature in the
core \citep{YLS99,YKGH01,YP04}.  Due to the large uncertainties in the
composition and physical state of the matter, we do not attempt to
model the superfluid core in detail.  Rather, to study the effects of
superfluidity on superburst characteristics, we simply fix the
temperature at the crust-core interface at several values and carry
out the calculations.  Figure 12 shows the results.  Neutron
stars with hotter cores exhibit superbursts with smaller fluences and
shorter recurrence times.  In addition, we see that for these
temperatures the lower limit of the accretion rate range over which
superbursts occur increases as the core temperature increases.  The
high crust temperature causes the carbon to burn stably before an
instability is triggered, in agreement with \citet{CB01}.

Impurities in the crust of a neutron star can significantly affect the 
thermal profile of the superburst ignition region, especially in stars with
cold cores (see $\S 3.5$).  Consequently, superburst energies and recurrence 
times are quite sensitive to the concentration of impurities in the crust.  
Figures 13 and 14 show some results.  We see that there is little 
difference between $Q = 0$ and $Q = 5.2$ when the core radiates 
neutrinos via modified URCA reactions, but $Q = 100$ induces a 
substantial change in the results.  However, recent investigations 
suggest that 
impurity scattering in the crusts of accreting neutron stars that 
exhibit superbursts is rather insignificant \citep{SBC03}.  If the 
impurity concentration $Q \sim 1$ in the crust of a neutron star with 
a core that emits neutrinos via pionic or direct URCA reactions, then the 
carbon fuel will solidify and burn stably via 
pycnonuclear reactions at lower accretion rates.  

We have shown in $\S 3.6$ that, for a completely disordered
neutron star crust, the thermal profile (including that of the superburst
ignition region) is highly insensitive to the core neutrino emission
mechanism.  In this case, superburst energies and recurrence times should 
be insensitive to the core cooling mechanism, as confirmed in Figure 15.  
Thus we verify the result of \citet{B04} that superburst
energies and recurrence times from neutron stars with highly efficient
core neutrino cooling mechanisms may still be consistent with
observations if the crust is disordered.  However, in this case we
find that superbursts should occur even at relatively low accretion rates
$\dot{M} < 0.1 \dot{M}_{\mathrm{Edd}}$, whereas observations indicate
a cutoff at $\dot{M} \approx 0.1 \dot{M}_{\mathrm{Edd}}$.  Normally, 
the carbon 
would burn stably at these low accretion rates because the energy generated 
would be efficiently transported away from the burning region.  However, the 
low thermal conductivity due to the disordered lattice inhibits the flow 
of energy away from the carbon-burning region.  
Therefore, even a low carbon energy generation rate can 
initiate a thermonuclear instability.  Note that in
deriving these results we assume that the entire crust is completely
disordered, which is clearly an extreme situation.  Further
investigations into the nuclear structure of neutron star crusts are
necessary to determine the significance of these results.

\subsection{Stellar Radius}

The radius of a neutron star depends quite sensitively on the core 
equation of state, but it is virtually independent of the stellar mass.  
Accurate measurements of the radius to within about one kilometer can 
potentially constrain the equation of state \citep{LP01}.  To demonstrate 
the effects of the stellar radius on superburst characteristics, we choose 
three different values, $R = 16.4$, $10.4$, and $6.5$ km, which likely 
bracket the true radii of neutron stars.  We find that 
stars with larger radii have more energetic superbursts and longer 
recurrence times at a given accretion rate.  They also have superbursts 
at lower accretion rates.  See Figure 16.  

At a given accretion rate, a neutron star with a larger radius
requires a larger column density of fuel in order for a superburst to
be triggered.  This is a result of the lower gravitational
acceleration near the stellar surface.  Thus the effect of radius on
superburst recurrence times is twofold.  Not only is the accretion
rate per unit area smaller for a larger star, but also the amount of
fuel per unit area required for a superburst to occur is larger.  The
effect of radius on superburst energetics is even stronger.  In
addition to the factors stated above, the total surface area of the
star is larger, so the total amount of fuel available to burn at a
given column depth is greater.  Figure 16 thus suggests that
superburst observations may be useful for constraining neutron star
radii.

\section{Comparison with Observations}

As noted in \S1, the nine superbursts that have been observed,
excluding those from GX 17+2, have integrated photon fluxes of $ \approx
10^{42}$ ergs.  The paucity of data makes the recurrence time of
superbursts difficult to determine, but observations imply a
recurrence time of $\sim 1$-$2$ years.  All nine superbursts occurred
in systems with accretion rates between $10\%$ and $30\%$ of the
Eddington limit.  Additionally, several superburst candidates have been 
observed in the near-Eddington accretor GX 17+2 with energies of 
$\sim 5 \times 10^{41}$ ergs.  A successful theoretical model of superbursts 
must explain these facts.

We begin by discussing the $\dot M$ range over which superbursts are
observed.  The lack of superbursts for accretion rates $\dot{M}
\lesssim 0.1 \dot{M}_{\mathrm{Edd}}$ arises naturally in our model and 
is rather simple to explain theoretically. As discussed in
\S4.3, at low $\dot M$, either the carbon fuel burns stably via
thermonuclear reactions before an instability is triggered (for high
core temperatures, Cumming \& Bildsten 2001) or the crust solidifies 
in the superburst ignition
region and the carbon burns stably via pycnonuclear reactions (for low
core temperatures).  In either case, there is a cutoff of superbursts
at low values of $\dot M$.

The above discussion is invalid if the crust is highly disordered,
since then the thermal conductivity is low, causing the region in
which superbursts are triggered to be hotter than if the crust were
ordered.  Also, when the carbon ignites, the low thermal conductivity
inhibits the diffusion of the nuclear energy generated there.  The
nuclear energy generation rate exceeds the rate at which thermal
conduction can cool the region, so a thermonuclear runaway ensues and
a superburst is triggered.  Therefore, as seen in Figure 15,
superbursts occur down to accretion rates below $0.1\dot M_{\rm
Edd}$.  The lack of observed superbursts for accretion rates $\dot{M}
\lesssim 0.1 \dot{M}_{\mathrm{Edd}}$ then needs to be explained.  One
possibility is that, for some reason, e.g., lack of delayed mixed 
bursts (see below), carbon is not produced in
sufficient quantities to fuel a superburst.  Another explanation could
be that at low accretion rates the recurrence times are so long that
astronomers simply have not observed any of the systems long enough to see
superbursts.

The absence of superbursts for $0.3 \dot{M}_{\mathrm{Edd}} \lesssim \dot{M} 
\lesssim 1.0\dot{M}_{\mathrm{Edd}}$ is not well understood.  Every theoretical
study, including our own, suggests that superbursts should occur more
easily at higher mass accretion rates, assuming a sufficient amount of
carbon is present.  The peak luminosities of superbursts are often
below the Eddington limit.  Therefore, one possible explanation is
that none has been observed in this range of accretion rates because there is
little contrast between the peak burst luminosity and the accretion
luminosity \citep{SB03}.  On the other hand, the recurrence time is
inversely proportional to the accretion rate to first order, so one
would expect to observe more superbursts at higher accretion rates.

Another possible explanation involves the rp-process.  Perhaps the
rp-process leaves behind enough unburnt carbon to produce superbursts
only for $\dot M < 0.3\dot M_{\rm Edd}$, and it burns too much carbon at
higher mass accretion rates.  Both the critical amount of carbon
needed for superbursts and the amount of unburnt carbon left by the
rp-process decrease with increasing $\dot M$.  Perhaps the two
dependencies conspire to permit superbursts only for $\dot M <0.3\dot
M_{\rm Edd}$.  One problem with this explanation is that all studies
of the rp-process carried out so far predict too little carbon to
produce superbursts at all $\dot M$.  

The above proposals do not explain the observed superbursts from GX
17+2, which accretes at $\dot{M} \sim \dot{M}_{\mathrm{Edd}}$.  If one 
can observe superbursts in a system accreting at nearly the Eddington 
limit, than presumably one could observe superbursts in systems accreting at 
any lower rate as well.  Furthermore, as
discussed in \S4.2, the minimum mass fraction of carbon needed to
trigger a superburst is inversely proportional to the accretion rate.
However, the amount of carbon left behind by stable burning via the
rp-process falls even more rapidly with increasing $\dot M$.

Yet another possible explanation involves ideas described in Paper I.
In that paper, Narayan and Heyl showed that at high accretion rates
($\dot{M} \gtrsim 0.1\dot{M}_{\mathrm{Edd}}$) normal Type I X-ray
bursts occur in a unique regime that they refer to as ``delayed mixed
bursts.''  In these systems, a large fraction of the hydrogen and
helium fuel burns stably to carbon before the instability is
triggered.  This stable burning explains the observations of
\citet{vPPL88} who found that, for systems that accrete at a high rate
and exhibit normal bursts, the quantity $\alpha$, the ratio of the
energy released between bursts to the energy released during a burst, 
rises dramatically.  Such an increase of $\alpha$ is
nicely reproduced by the model described in Paper I.  Furthermore,
\citet{intZetal03} showed that systems that exhibit superbursts
generally have $\alpha$ values that are significantly greater than
systems that do not exhibit superbursts.

Delayed mixed bursts may be the source of the substantial amount of
carbon needed to trigger a superburst.  As noted earlier, every system 
in which a superburst
has been observed exhibits normal bursts as well.  We tentatively
suggest that the occurrence of delayed mixed bursts is a necessary
condition for superbursts in systems for which the accreted material
is predominantly hydrogen.  The stably burned material forms a thick
layer of carbon, and the delayed burst that finally occurs burns above
this layer, leaving the carbon largely unaffected.  This would explain
the absence of observed superbursts in systems with accretion rates
$0.3 \dot{M}_{\mathrm{Edd}} \lesssim \dot{M} 
\lesssim 1.0\dot{M}_{\mathrm{Edd}}$.
These systems do not exhibit normal Type I bursts, so the hydrogen and helium
fuel burns stably via the rp-process.  According to estimates in the
literature, the rp-process leaves behind too little carbon for a
superburst.  Note that the existence of normal bursts in GX 17+2 is not 
understood theoretically if the accreted material is indeed predominantly 
hydrogen.  However, \citet{Ketal02} find that $\alpha \gtrsim 1000$ in this 
system, which 
implies that the normal bursts are delayed bursts.  We emphasize that this 
explanation is merely a
hypothesis, and it is not without issues.  In particular, we do not
know if the carbon produced during the stable burning phase of a
delayed burst system would survive when the burst is triggered above
it.  More research is needed on this interesting problem.

Delayed mixed bursts may also explain the absence of superbursts for
$\dot{M} < 0.1 \dot{M}_{\mathrm{Edd}}$ in models with disordered
crusts.  These models predict superbursts down to relatively low accretion
rates, provided they have enough carbon.  However, Paper I showed that
at accretion rates below $0.1\dot M_{\rm Edd}$ the normal Type I
bursts are prompt bursts.  Such bursts are expected
to burn all the fuel to elements much heavier than carbon, and so
their ashes should not support superbursts.

We next consider the energies and recurrence times of superbursts.
Leaving aside models with disordered crusts, Figure 11 shows
that neutron stars with ordered crusts and highly efficient neutrino
emission mechanisms in their cores have much more powerful superbursts
and longer recurrence times than are observed.  Furthermore, if the 
impurity parameter $Q \sim 1$, these
stars do not exhibit superbursts at accretion rates $\dot{M} \lesssim
0.3 \dot{M}_{\mathrm{Edd}}$, where 
all superbursts except those from GX 17+2 have been observed 
(see Fig.\ 14).  Therefore, we conclude that accreting
neutron stars with highly efficient neutrino emission in their cores
(due to direct URCA or pionic reactions, for example) are inconsistent
with superburst observations, in agreement with \citet{B04}.

Even for neutron stars that emit neutrinos via modified URCA reactions
in their cores, we predict energies and recurrence times somewhat
larger than those observed.  The observed superburst energies are
$\sim 10^{42}$ ergs, which is roughly an order of magnitude lower than
those in Figure 11 for modified URCA cooling.  However, a substantial
fraction of the superburst energy may be released in the form of
neutrinos \citep{SB02}, so this discrepancy may not be serious.
Potentially more troublesome are the relatively long superburst
recurrence times predicted by the model.  Although observations do not
constrain recurrence times very well, the times are probably shorter 
than those in Figure 11.  Neutron star models with hot cores,
$T_{\mathrm{core}} \gtrsim 5 \times 10^{8}$ K, seem to best match the
observational data, though the $\dot{M}$ cutoff is 
larger (see Figure 12).  Such temperatures
correspond to neutrino cooling that is even less efficient than that 
from modified URCA reactions.  One
possibility is that a nonnegligible fraction of the neutron star core 
consists of superfluid baryonic matter.

The above conclusions about the neutrino emission mechanism in the
core assume that the stellar radius is the canonical $10$ km.  The
radius of the neutron star is the only other parameter which we have
investigated that significantly affects superburst characteristics at
the accretion rates at which superbursts are observed.  Larger stars
produce superbursts with larger energetics and longer recurrence times
(see Figure 16).  Neutron stars with exceptionally large radii
($R \approx 16.4$ km) produce extremely energetic superbursts that are
grossly inconsistent with observations, even if the core temperature
is very high.  We cannot make such a definitive statement about
neutron stars with smaller radii because we are unable to
differentiate between superbursts from neutron stars with small radii
($R \approx 6.5$ km) and moderate core temperatures
($T_{\mathrm{core}} \approx 3 \times 10^{8}$ K) and neutron stars with
moderate radii ($R \approx 10.4$ km) and high core temperatures
($T_{\mathrm{core}} \approx 8 \times 10^{8}$ K).

\section{Comparison with Previous Theoretical Work}

Previous theoretical investigations of the superburst phenomenon 
(Brown \& Bildsten 1998, Cumming \& 
Bildsten 2001, Strohmayer \& Brown 2002, Brown 2004) generally agree 
quite well with 
our results.  To determine the physical conditions under which superbursts 
occur, the authors use approximate ignition criteria evaluated at 
the base of the accreted layer.
In this section, we compare and contrast the results of our rigorous 
global linear stability analysis with those obtained from these approximate 
one-zone ignition criteria.

According to \citet{BB98}, an instability ensues when the carbon nuclear 
energy generation rate $\epsilon_{\mathrm{C}}$ at the base of the accreted 
layer satisfies the criterion
\begin{equation}
\frac{\mathrm{d} \epsilon_{\mathrm{C}}}{\mathrm{d} T} > 
\frac{\mathrm{d} \epsilon_{\mathrm{cool}}}{\mathrm{d} T}, 
\end{equation}
where $\epsilon_{\mathrm{cool}} = \rho K T / \Sigma^{2}$ is an 
approximation to the global cooling rate and $K$ is the thermal 
conductivity evaluated at the base of the accreted layer.  \citet{CB01} 
and \citet{B04} set $\mathrm{d} \ln \epsilon_{\mathrm{C}} / 
\mathrm{d} \ln T = 26$ and $\mathrm{d} \ln \epsilon_{\mathrm{cool}} / 
\mathrm{d} \ln T = 2$ and express the ignition criterion as
$\epsilon_{\mathrm{C}} > (2/26) \epsilon_{\mathrm{cool}}$.  We presume 
that \citet{BB98} and \citet{SB02} calculate the derivatives numerically.  
Unlike 
\citet{BB98} and \citet{SB02}, \citet{CB01} and \citet{B04} do not 
explicitly state that they track the evolution of the carbon mass 
fraction $Z_{\mathrm{CNO}}$ when they solve for the equilibrium 
configuration of the layer.  To our knowledge, these authors keep 
$Z_{\mathrm{CNO}}$ constant throughout the layer, but we cannot state 
this with certainty.

To conduct an accurate comparison between our global analysis and the 
various one-zone approximations, we solve for the equilibrium configuration 
of the accreted layer in an identical fashion for both methods.  This 
ensures that any differences in the results are due only to the stability 
calculation.  We find that $\mathrm{d} \ln \epsilon_{\mathrm{cool}} / 
\mathrm{d} \ln T \approx 2$ in all of our calculations.  Therefore, 
setting $\mathrm{d} \ln \epsilon_{\mathrm{cool}} / 
\mathrm{d} \ln T = 2$ is appropriate for all scenarios in our opinion.  
In contrast, 
we find that setting $\mathrm{d} \ln \epsilon_{\mathrm{C}} / 
\mathrm{d} \ln T = 26$ is appropriate only if the core temperature is 
high ($\gtrsim 10^{8} $K) or the accretion rate is near the Eddington 
limit.  At sufficiently high densities and low temperatures, 
$\mathrm{d} \ln \epsilon_{\mathrm{C}} / \mathrm{d} \ln T \ll 26$, often 
by many orders of magnitude, so this 
approximation is inappropriate in these situations.  

For a given equilibrium configuration, we calculate the one-zone 
approximation in four different ways.  When calculating 
$\mathrm{d} \epsilon_{\mathrm{C}} / \mathrm{d} T$ at the 
base of the accreted layer we (i) 
use the value of $Z_{\mathrm{CNO}}$ and $\mathrm{d} \ln 
\epsilon_{\mathrm{C}} / 
\mathrm{d} \ln T$ derived from our equilibrium configuration, (ii) 
artificially set $Z_{\mathrm{CNO}}$ constant and use the value of
$\mathrm{d} \ln \epsilon_{\mathrm{C}} / 
\mathrm{d} \ln T$ derived from our equilibrium configuration, (iii) 
use the value of $Z_{\mathrm{CNO}}$ derived from 
our equilibrium configuration and set $\mathrm{d} \ln \epsilon_{\mathrm{C}} 
/ \mathrm{d} \ln T = 26$, or (iv) artificially set $Z_{\mathrm{CNO}}$ 
constant and
set $\mathrm{d} \ln \epsilon_{\mathrm{C}} / \mathrm{d} \ln T = 26$.  
Artificially setting $Z_{\mathrm{CNO}}$ constant results in superbursts 
at all accretion rates for each calculation we performed.  
This is clearly incorrect since all our global stability analysis 
models indicate a minimum 
$\dot{M}$ below which superbursts are absent.  Therefore, the 
criteria (ii) and (iv) above are very inaccurate and should be 
avoided.  Regarding the other two criteria, we find that for neutron 
stars with high core temperatures, such as those with cores that 
radiate neutrinos via modified URCA reactions, 
the results from the calculations in which 
$\mathrm{d} \ln \epsilon_{\mathrm{C}} / \mathrm{d} \ln T$ is set 
to $26$ and in which 
$\mathrm{d} \ln \epsilon_{\mathrm{C}} / \mathrm{d} \ln T$ is calculated 
self-consistently are almost identical.  Therefore, the approximation 
$\mathrm{d} \ln \epsilon_{\mathrm{C}} / \mathrm{d} \ln T = 26$ is 
valid in this situation.  However, for neutron stars with low core 
temperatures, such as those with cores that radiate neutrinos via pionic 
reactions, the results from the two calculations differ significantly, so 
in this case $\mathrm{d} \ln \epsilon_{\mathrm{C}} / \mathrm{d} \ln T$ 
must be calculated self-consistently.  

See Figures 17 and 18 for a comparison between the results from our 
global linear stability analysis and the one-zone approximations (i) 
and (iii).  For the 
calculations in which both $Z_{\mathrm{CNO}}$ and 
$\mathrm{d} \ln \epsilon_{\mathrm{C}} / \mathrm{d} \ln T$ are calculated 
self-consistently, the results of the one-zone approximation and the 
global linear stability analysis generally agree quite well.  The 
errors produced by the one-zone analysis are probably smaller than those 
due to uncertainties in other parameters, such as the accretion rate, 
impurity concentration, elemental composition of the crust, etc.  The 
only significant discrepancy we find between the two models is the 
superburst recurrence times in systems with accretion rates near the 
critical rate below which superbursts do not occur.  In these systems, 
we find from our global linear stability analysis that the carbon layer 
is marginally unstable.  
A large fraction of the carbon fuel burns stably before the full instability 
is triggered, resulting in a delayed burst (see Paper I), whereas the 
one-zone calculation has no way of modeling delayed bursts.  These 
``delayed'' superbursts have significantly larger recurrence times than 
those derived using the one-zone approximation.  However, they occur 
over a fairly 
narrow range of accretion rates, so this discrepancy may not be serious.  
From our comparison, we draw the following conclusions regarding the 
application of the one-zone approximation to superburst calculations:
\textit{The one-zone approximation is appropriate for most superburst 
calculations if one uses the values of $Z_{\mathrm{CNO}}$ and 
$\mathrm{d} \ln 
\epsilon_{\mathrm{C}} / \mathrm{d} \ln T$ derived from the equilibrium 
configuration calculation.  The resulting superburst energies should be 
sufficiently accurate for all accretion rates.  The superburst recurrence 
times should be sufficiently accurate at high accretion rates, but they 
will significantly underestimate the true recurrence times at accretion 
rates near the critical rate.  All simpler approximations, such as 
(ii), (iii), and (iv) above, make serious errors for some choices of 
parameters.}

\section{Summary}

In this investigation, we have carried out the first self-consistent
global linear stability analysis of the carbon fuel on accreting
neutron stars and determined the physical conditions under which
superbursts occur.  Our model reproduces the general observed
superburst features, including burst energies, recurrence times, and
the range of accretion rates at which superbursts occur.  In contrast 
to normal Type I X-ray bursts, the observational characteristics of
superbursts are very sensitive to the thermal profile of the entire
crust of the neutron star.  Consequently, superbursts can be useful
probes to study neutron star interiors.  Because our theoretical model
evaluates the thermal profile self-consistently, we are able to
explore a wide range of neutron star parameters and study the effects
of each on the resulting superburst energies and recurrence times.

By comparing our results with observations, we are able to set
constraints on various neutron star parameters.  We find that
accreting neutron stars with highly efficient neutrino emission in
their cores produce extremely energetic superbursts which are
inconsistent with observations. Such neutron stars also do not have 
superbursts in
the range of accretion rates at which superbursts are observed unless 
the crust is very impure.  Stars
with less efficient neutrino emission produce bursts that agree better
with observations, while stars with highly suppressed neutrino
emission in their cores, e.g., because of superfluidity, produce
bursts that agree best with observations.  

If the neutron star crust is disordered, the thermal profile of the
crust, in particular the region in which superbursts are triggered, is
insensitive to the core temperature.  Therefore, superburst energetics
and recurrence times from neutron stars with highly efficient core
neutrino cooling mechanisms may still be consistent with observations
if the crust is completely disordered.  However, such neutron stars 
will have superbursts at accretion rates lower than the observed cutoff 
rate.  The cutoff at $\dot{M} \sim 0.1 \dot{M}_{\mathrm{Edd}}$ thus 
requires some other effect, e.g.\ delayed mixed bursts.

Neutron stars with large radii ($R \sim 16$ km) produce very energetic
superbursts that are inconsistent with observations, even if the core
neutrino emission mechanism is highly inefficient.  Better constraints
on neutron star parameters could be made with improvements to the
theoretical model and with more observational data.

All systems in which superbursts are observed and that accrete predominantly 
hydrogen have exceptionally high $\alpha$ values (see \S5).
Delayed mixed bursts, which are normal bursts that burn a significant
amount of hydrogen and helium stably to carbon before the instability
is triggered (Paper I), occur in systems for which $0.1 \lesssim
\dot{M}/\dot{M}_{\mathrm{Edd} } \lesssim 0.3$ and produce $\alpha$
values consistent with those observed in these systems.  Since the
delayed mixed burst instability occurs above a layer of carbon that
has been already produced by stable burning, it is possible that they
leave behind sufficient amounts of unburnt carbon to fuel superbursts.
For systems in which delayed mixed bursts do not occur,
it is likely that the rp-process burns nearly all the available fuel
to elements heavier than carbon, leaving insufficient carbon for
superbursts.  We thus speculate that delayed mixed bursts are a necessary 
prerequisite for the occurrence of superbursts in systems that accrete 
predominantly hydrogen.  More work is needed to verify this suggestion.

Previous theoretical investigations have used approximate 
one-zone ignition criteria to determine the conditions under which 
superburst occur.  We find that the results of these approximate 
criteria generally agree well with our global stability analysis if 
the parameters used in the one-zone criteria are consistent with those 
derived in the equilibrium configuration calculation.  Specifically, 
calculations in which the carbon mass fraction is kept constant do not 
reproduce the absence of superbursts at lower accretion rates, and 
calculations which set $\mathrm{d} \ln \epsilon_{\mathrm{C}} / 
\mathrm{d} \ln T = 26$ produce inaccurate results for neutron 
stars with highly efficient neutrino emission mechanisms in their cores.  
For calculations in which these quantities are consistent with those 
derived in the equilibrium configuration calculation, the only 
significant discrepancy we 
have found is for systems that accrete at a rate close to the critical 
accretion rate below which superbursts do not occur.  At these 
accretion rates, 
a substantial amount of carbon burns stably before a superburst is 
triggered.  Consequently, the one-zone approximation drastically 
underestimates the superburst recurrence times.
 
\acknowledgments

It is our pleasure to thank Edward Brown, Jean in't Zand, and Don Lamb 
for helpful discussions and the referee for insightful comments and 
suggestions.  This work was supported by NASA grant NNG04GL38G.


\newpage

\begin{figure}
\plotone{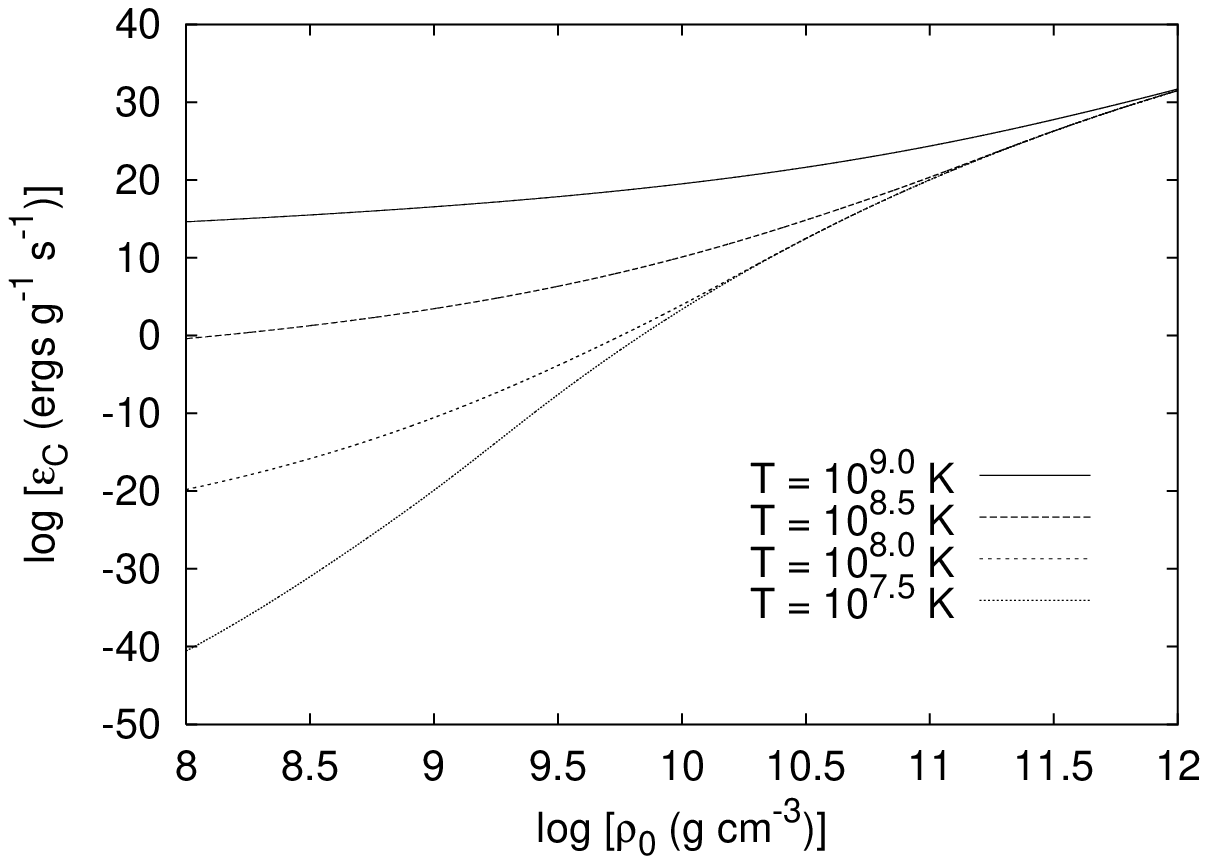}
\caption{Carbon energy generation rates as a function of rest mass density, 
plotted for four different temperatures.  The composition assumed 
is $30\%$ $^{12}$C and $70\%$ $^{56}$Fe by mass.  Note that at 
high densities, when the burning is strictly pycnonuclear, the energy 
generation rate is essentially independent of temperature. 
}
\label{fig1}
\end{figure}

\begin{figure}
\plottwo{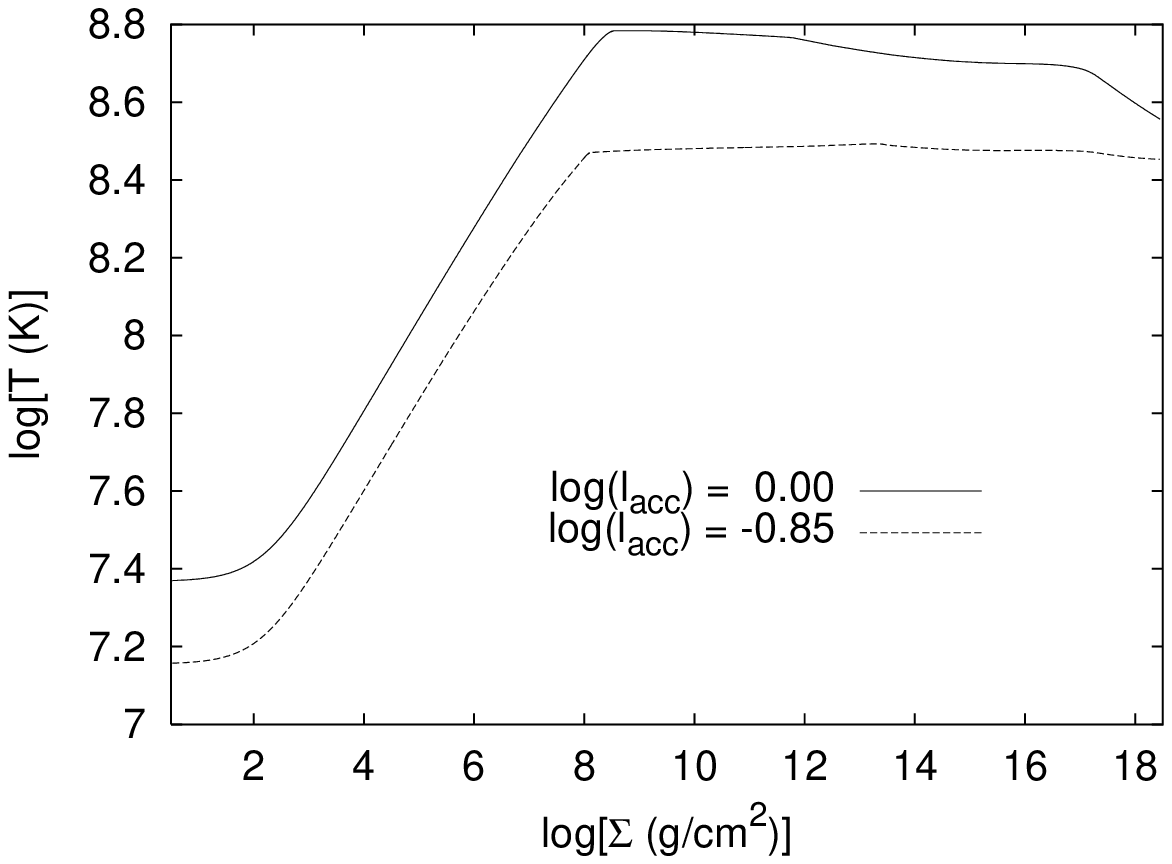}{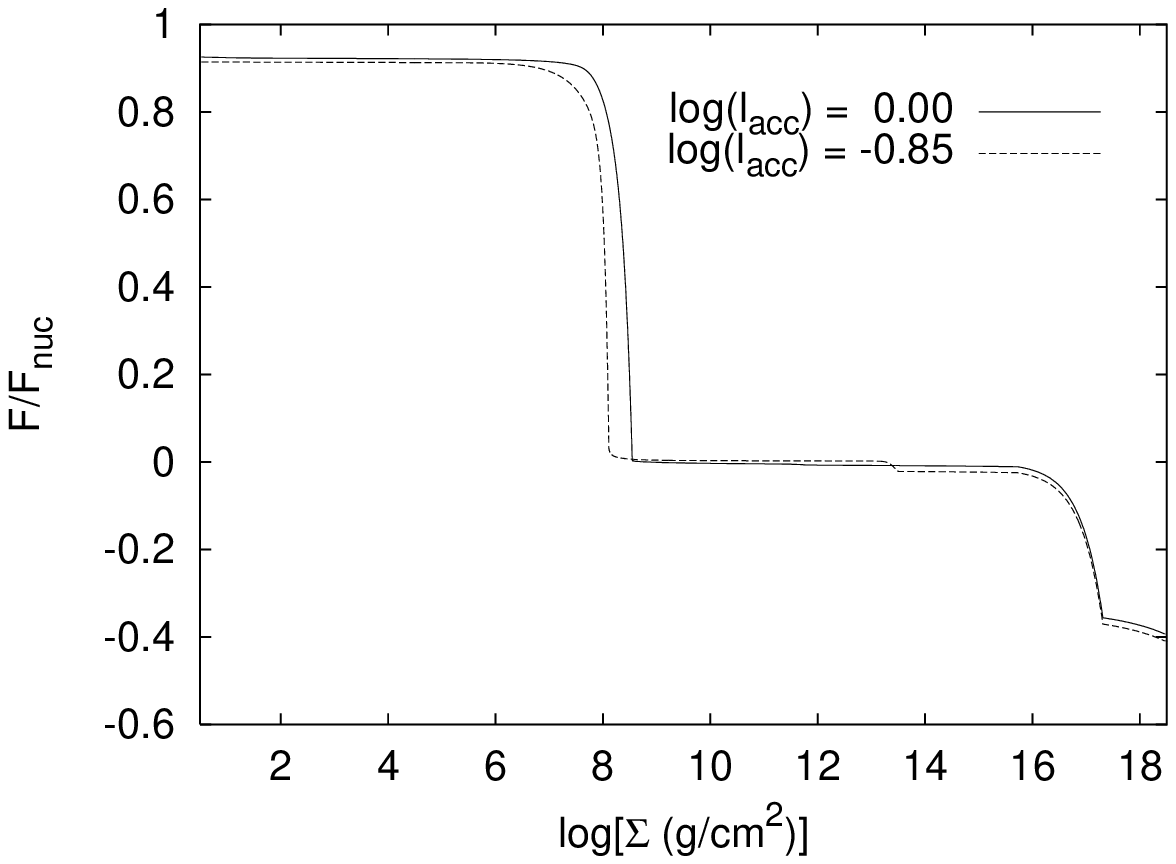}
\caption{Thermal and net energy flux profiles for two neutron stars with 
different mass accretion rates just before a superburst is triggered.  The 
parameter $l_{\mathrm{acc}}$ is the accretion rate normalized to the 
Eddington limit.  The rate of accretion of matter onto the surface of a 
neutron star significantly affects the thermal state of the entire crust, 
as well as the stellar core.  For this plot we assume fiducial parameter 
values $M = 1.4$ $M_{\odot}$, 
$R = 10.4$ km, $C_{\mathrm{f}} = 0.3$, $X_{\mathrm{out}} = 0.7$, 
$Y_{\mathrm{out}} = 0.28$, and $Z_{\mathrm{CNO,out}} = 0.016$.  
Additionally, we assume that 
the heavy element in the accreted layer is $^{56}$Fe and that the core 
emits neutrinos via modified URCA reactions.  These assumptions hold for 
all other Figures unless notified otherwise.
}
\label{fig2}
\end{figure}

\begin{figure}
\plotone{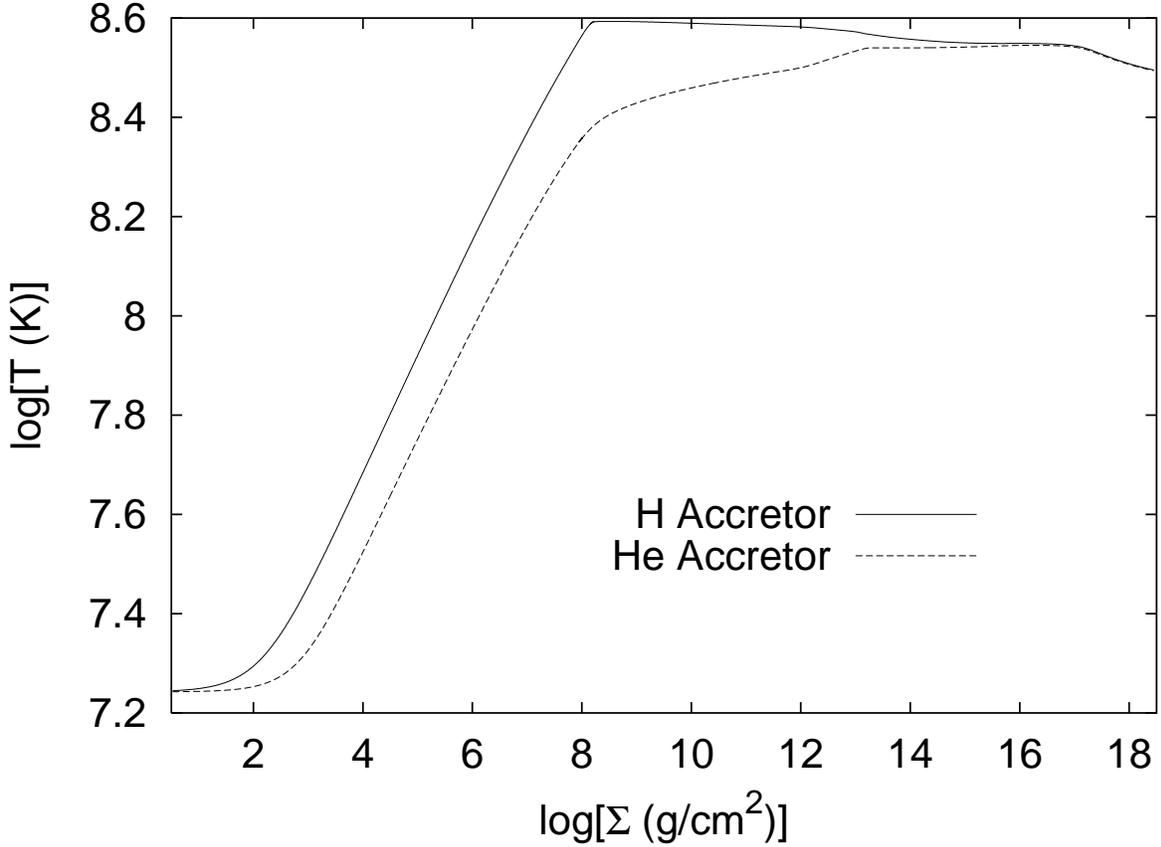}
\caption{Thermal profiles for two neutron stars with different accreted 
gas compositions just before a superburst is triggered.  The normalized 
accretion rate $l_{\mathrm{acc}} = 0.3$ for this and all subsequent 
thermal profiles.  The mass fractions of the accreted gas are  $X = 0.7$, 
$Y = 0.28$, $Z = 0.004$, and $Z_{\mathrm{CNO}} = 0.016$ for ``H Accretor,'' 
and $X = 0.1$, $Y = 0.88$, $Z = 0.004$, and $Z_{\mathrm{CNO}} = 0.016$ 
for ``He Accretor.''  Hydrogen burning releases much more energy per 
gram of fuel than helium burning.  Consequently, the temperature of the 
superburst ignition region, 
$10^{11}$ $\mathrm{g\,cm}^{-2} \lesssim \Sigma \lesssim 10^{14}$ 
$\mathrm{g\,cm}^{-2}$, 
is greater for hydrogen accretors than for helium accretors. 
}
\label{fig3}
\end{figure}

\begin{figure}
\plotone{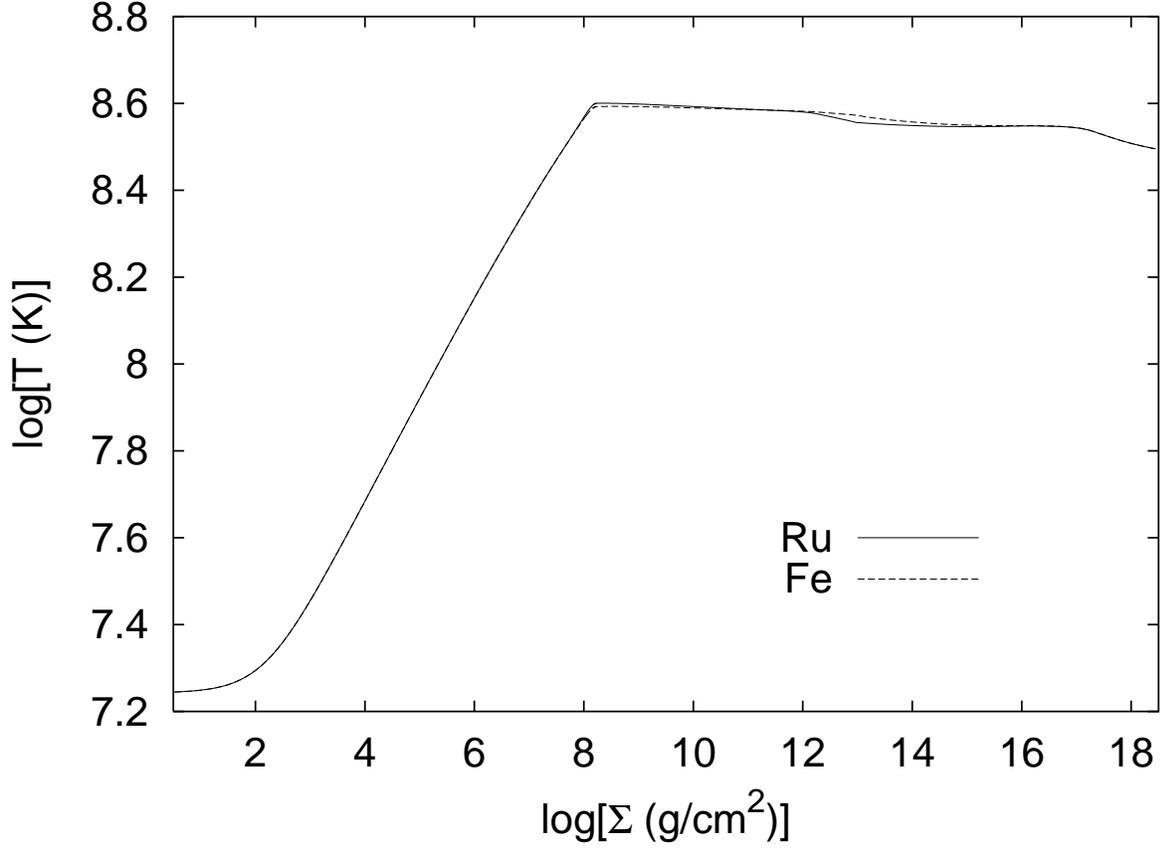}
\caption{Thermal profiles for two neutron stars with different heavy element 
compositions in the accreted layer resulting from stable and 
unstable hydrogen and helium burning just before a superburst is triggered. 
``Ru'' signifies a composition 
consisting of $30\%$ $^{12}$C and $70\%$ $^{104}$Ru by mass, and ``Fe'' 
signifies a composition 
consisting of $30\%$ $^{12}$C and $70\%$ $^{56}$Fe by mass.
The heavy element composition has almost no effect on the thermal profile 
of the neutron star crust.
}
\label{fig4}
\end{figure}

\begin{figure}
\plottwo{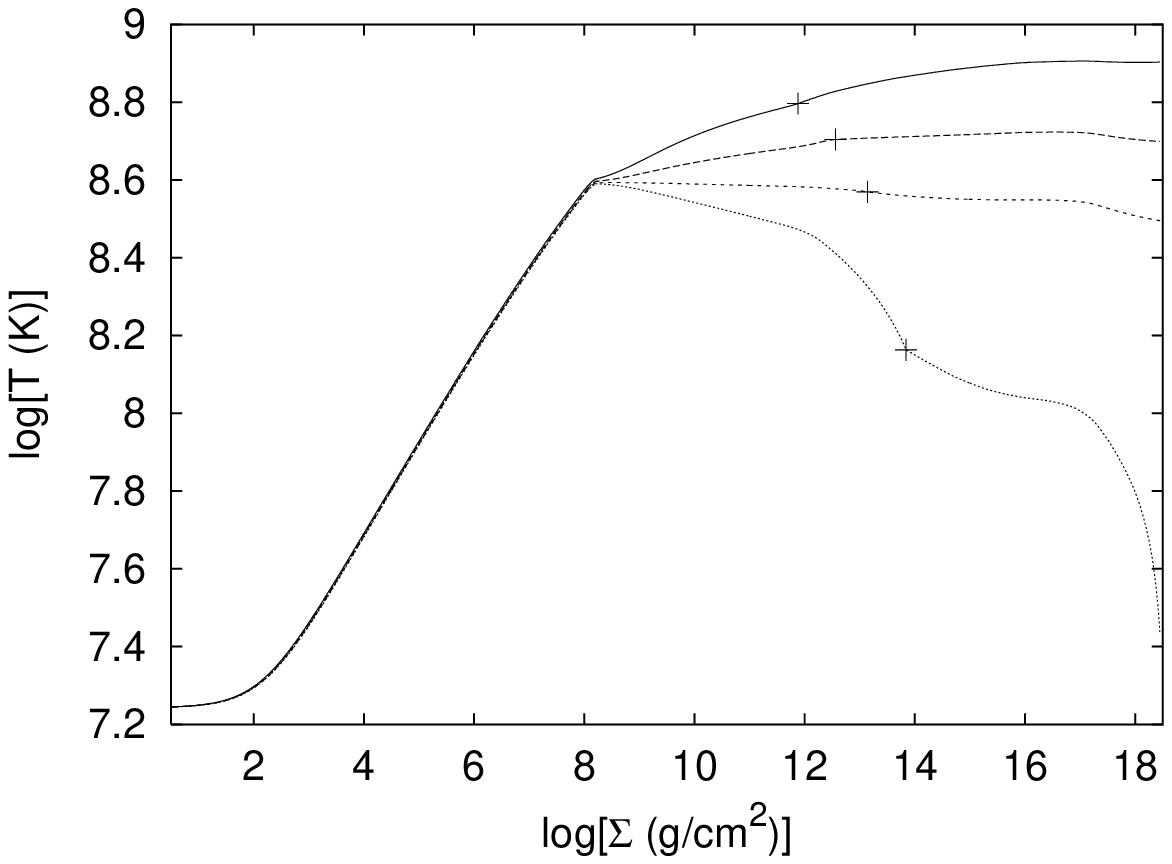}{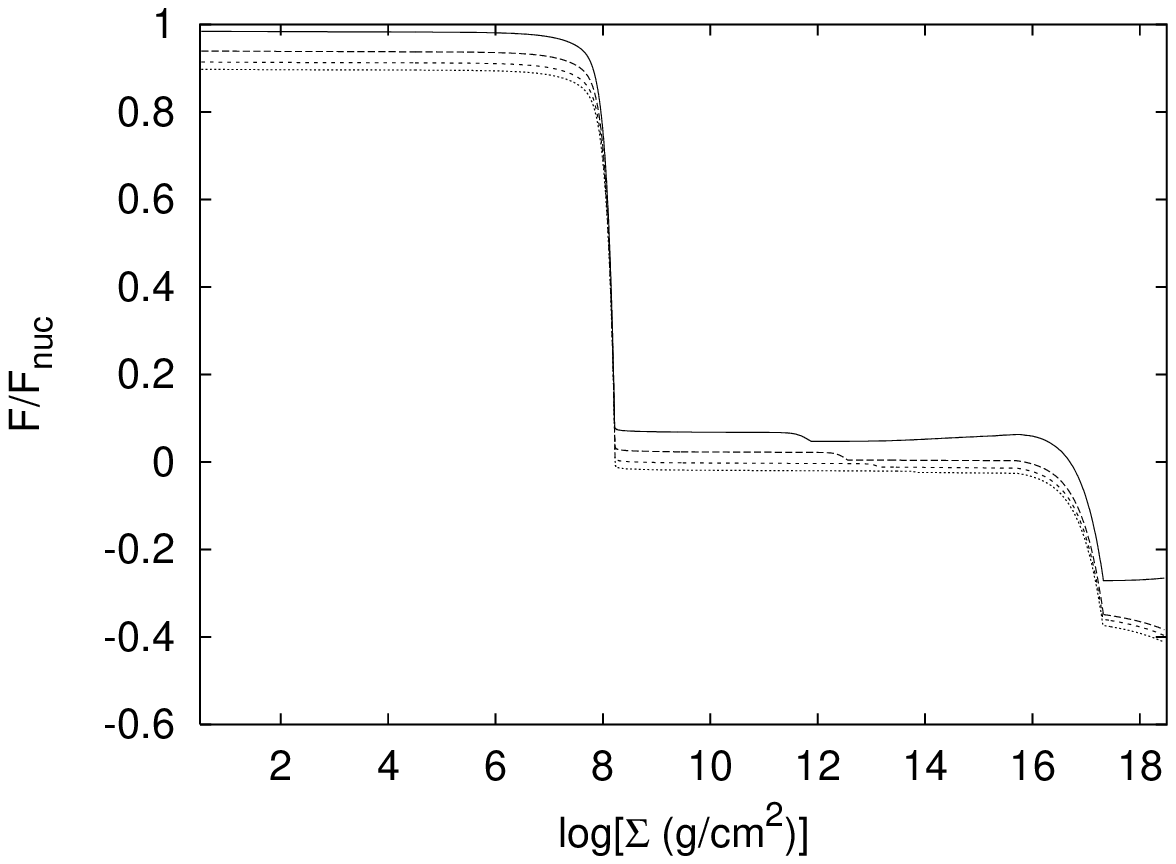}
\caption{Thermal and net energy flux profiles of the crust for neutron 
stars with different core temperatures just before a superburst is 
triggered.  Note that a positive value of $F$ 
denotes an outward flux.  The crosses indicate the column depth at which 
a superburst is triggered.  The core temperature significantly affects the 
thermal profile of the superburst ignition
region, $10^{11}$ $\mathrm{g\,cm}^{-2} \lesssim \Sigma \lesssim 10^{14}$ 
$\mathrm{g\,cm}^{-2}$.  
The flux near this region is sensitive to several parameters including 
the core temperature, and it can be either positive or negative.
}
\label{fig5}
\end{figure}

\begin{figure}
\plotone{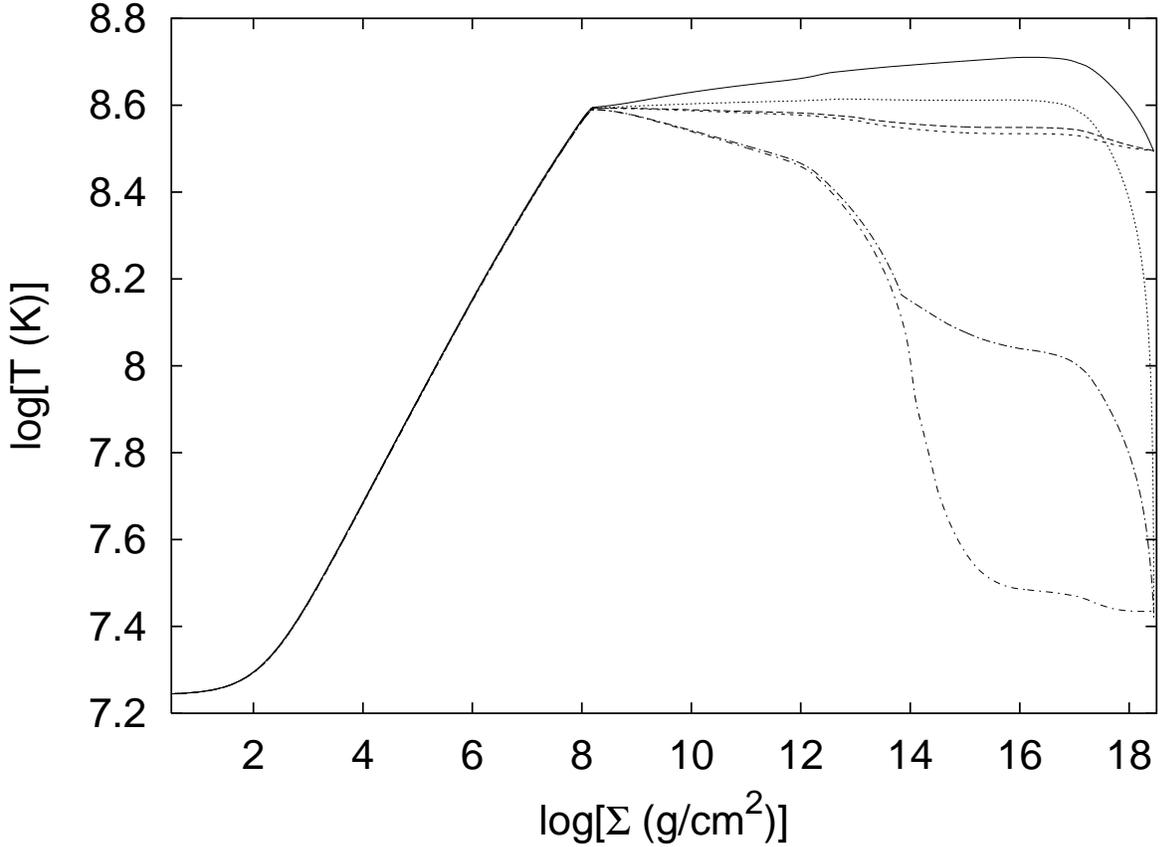}
\caption{Thermal profiles for accreting neutron stars with different values 
of the impurity parameter in the substrate, below the accreted layer.  We 
plot three thermal profiles for a neutron star with a core that emits 
neutrinos via modified URCA reactions (curves with $\log T \sim 8.5$ at 
$\log \Sigma \sim 18$) and three profiles for a star with 
a core that emits neutrinos via pionic reactions (curves with $\log T 
\sim 7.4$ at $\log \Sigma \sim 18$).  For each star with a 
given core cooling mechanism, the three profiles correspond to 
impurity parameter values $Q = 100$, $5.2$, and $0$ from hottest to 
coldest.  Note that impurity scattering has a significant effect on the 
thermal profile of the crust, especially in neutron stars with cold cores 
for which phonon scattering is weak.
}
\label{fig6}
\end{figure}

\begin{figure}
\plotone{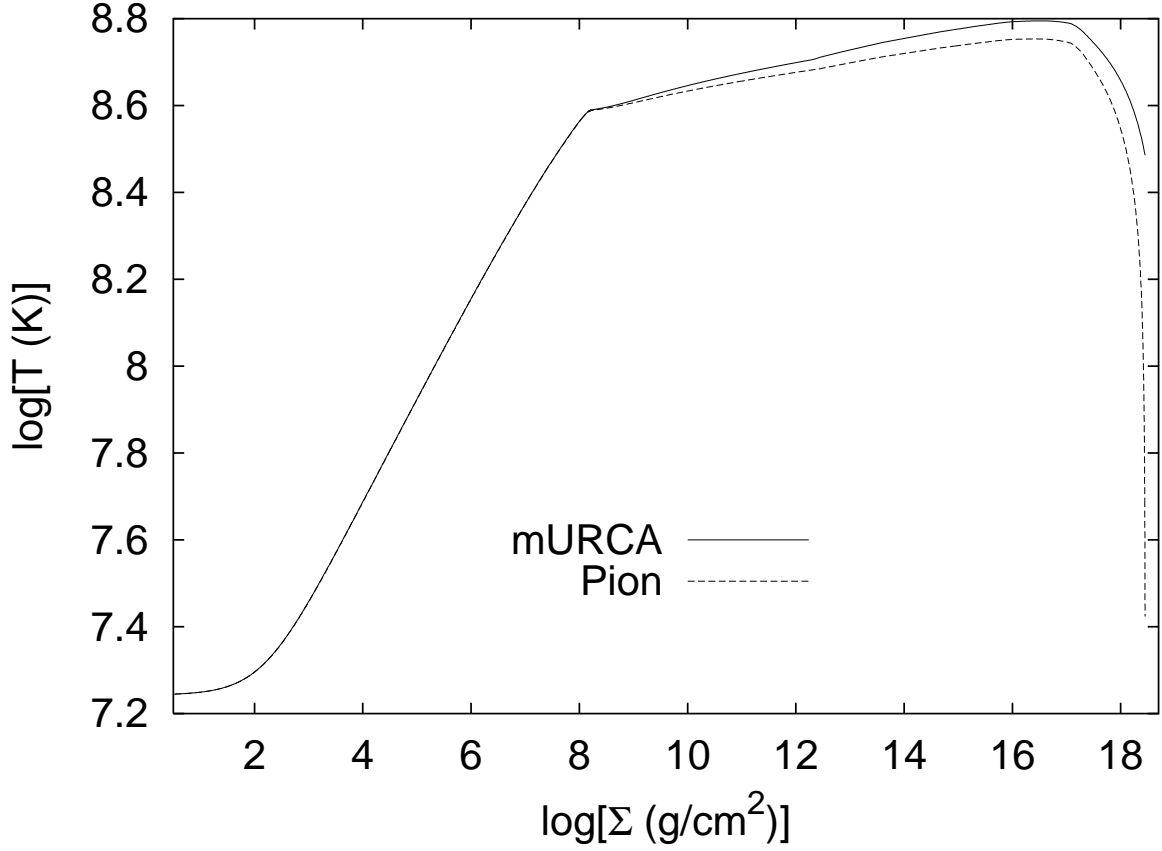}
\caption{Thermal profiles for accreting neutron stars with disordered 
crusts.  ``mURCA'' refers to a neutron star with a nonsuperfluid core that 
emits neutrinos via modified URCA reactions, and ``Pion'' refers to a 
neutron star with a core that emits neutrinos via pionic reactions.  The 
low thermal conductivity of the disordered crust makes the crust rather 
insensitive to the core temperature.  Contrast these results with 
those shown in Fig.\ 5.
}
\label{fig7}
\end{figure}

\begin{figure}
\plottwo{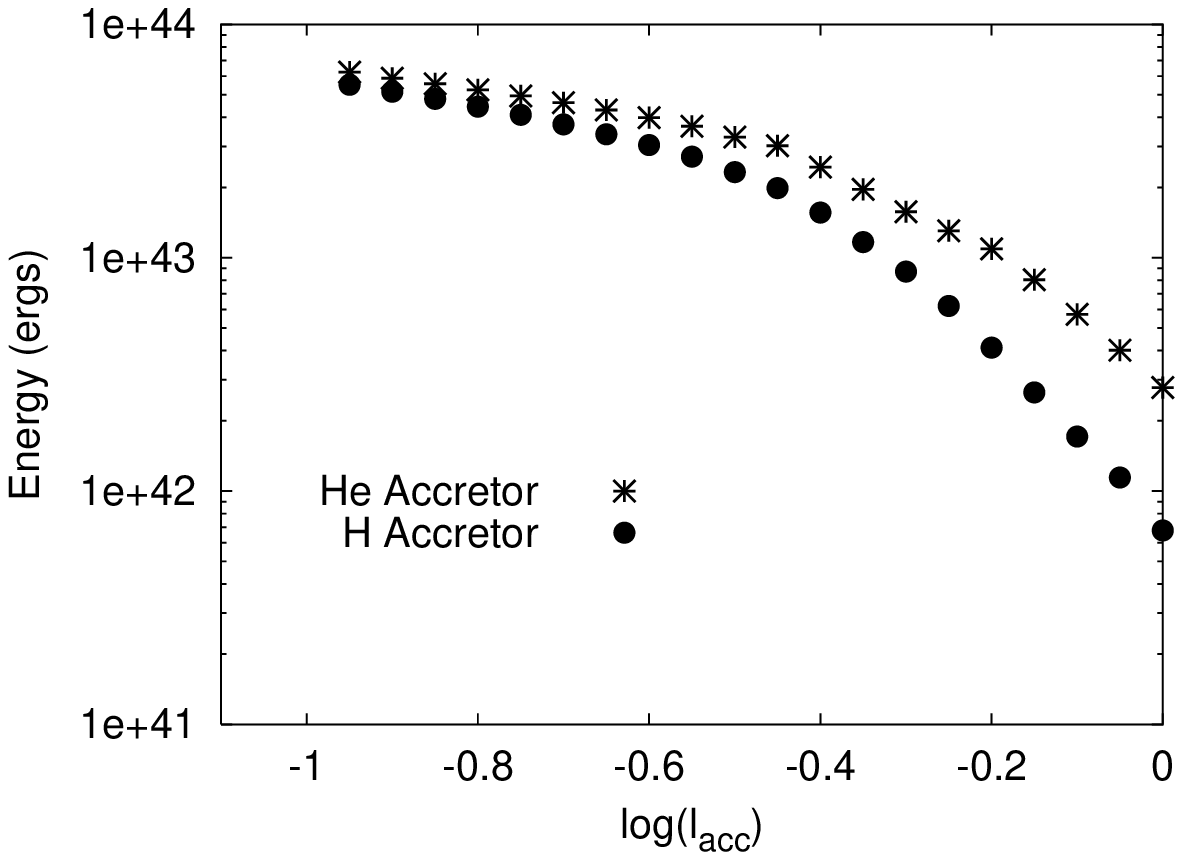}{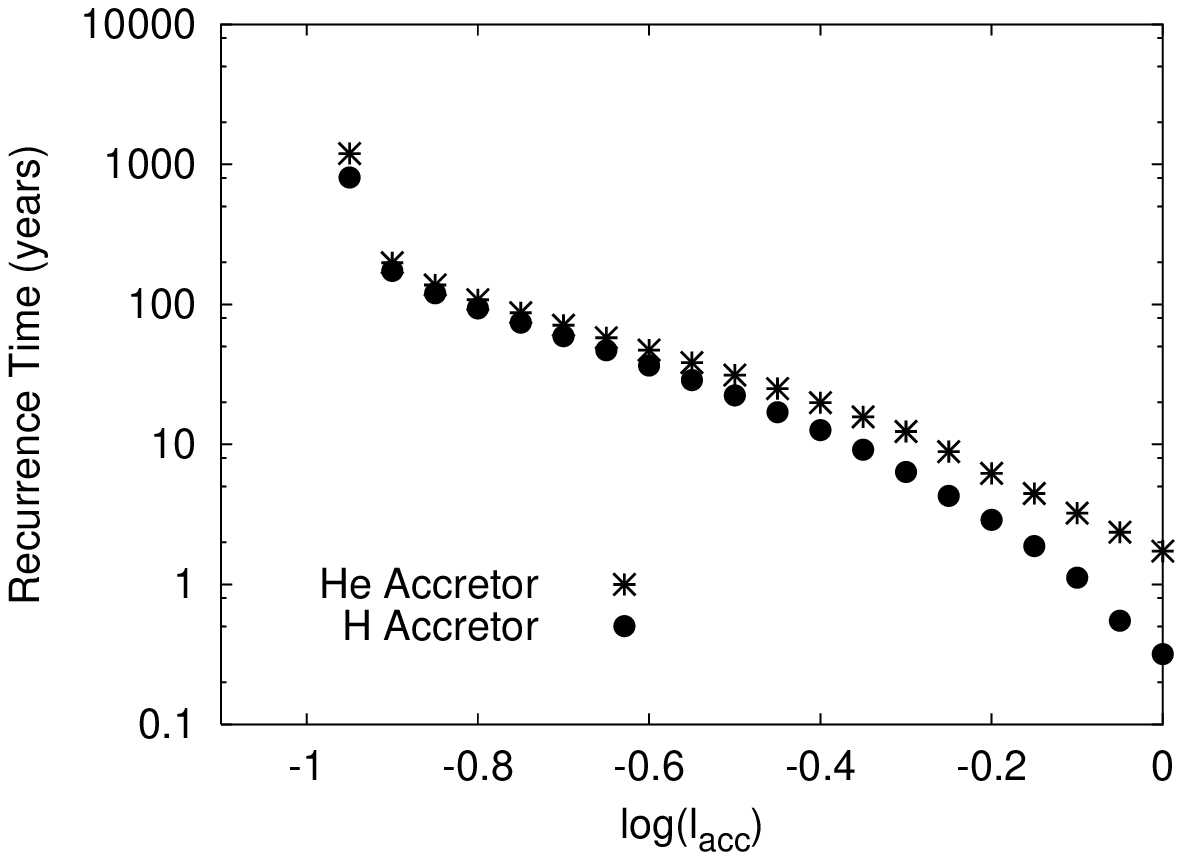}
\caption{Superburst energies and recurrence times for mixed 
hydrogen/helium accretors and for 
helium accretors as a function of accretion rate.  The core is 
assumed to consist of normal (nonsuperfluid) matter and to emit neutrinos 
via modified URCA reactions.  The composition of the burned hydrogen 
and helium is taken to be $30\%$ $^{12}$C and 
$70\%$ $^{56}$Fe by mass (i.e.\ $C_{\mathrm{f}} = 0.3$).  We assume the 
fiducial value $C_{\mathrm{f}} = 0.3$ for all calculations unless 
notified otherwise.
}
\label{fig8}
\end{figure}

\begin{figure}
\plottwo{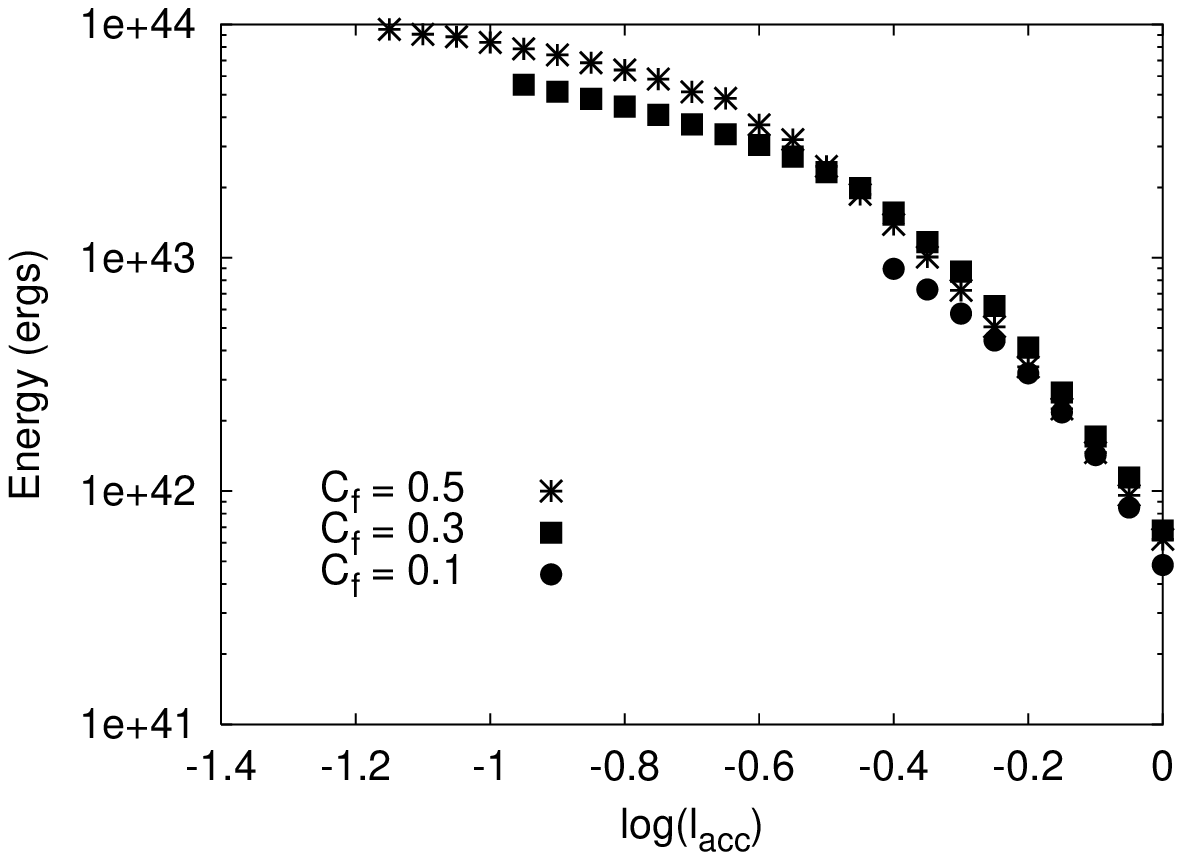}{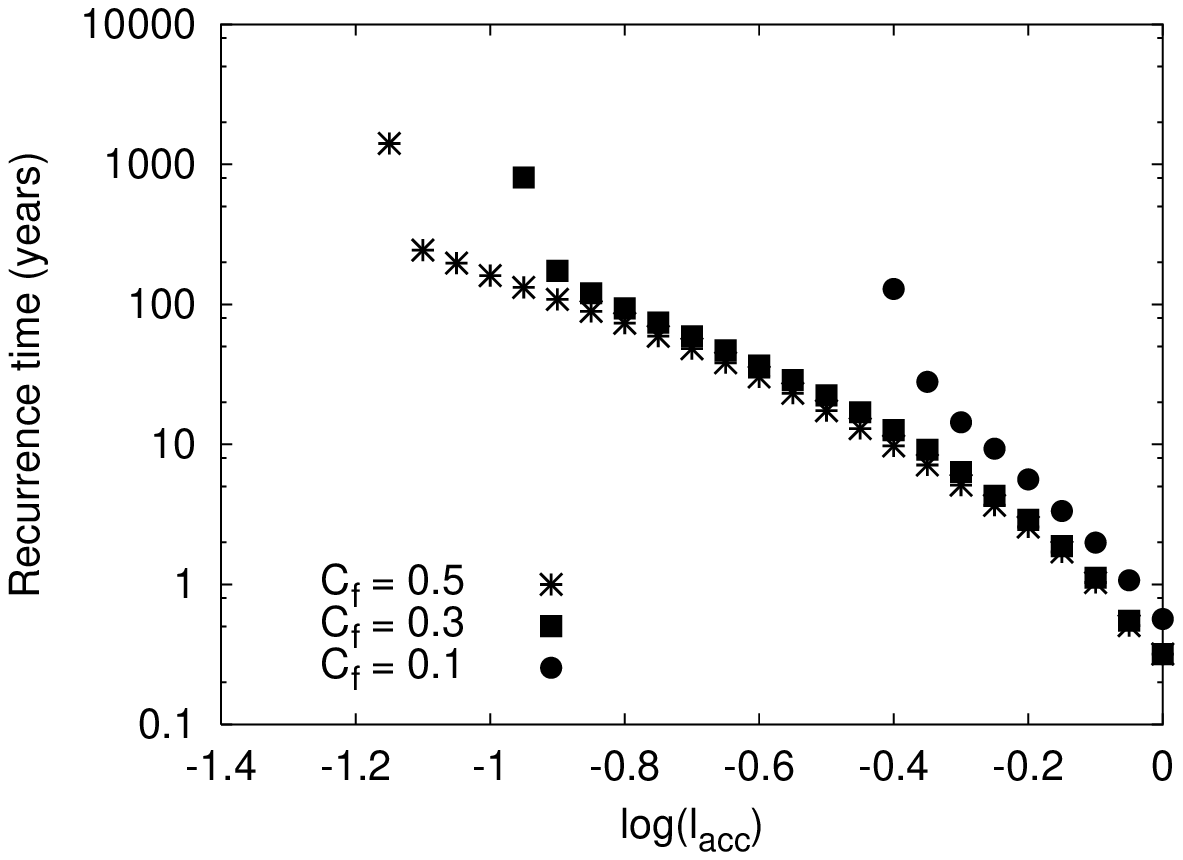}
\caption{Superburst energies and recurrence times as a function of 
accretion rate for different values of $C_{\mathrm{f}}$.  The core is 
assumed to emit neutrinos via modified URCA reactions.
}
\label{fig9}
\end{figure}

\begin{figure}
\plottwo{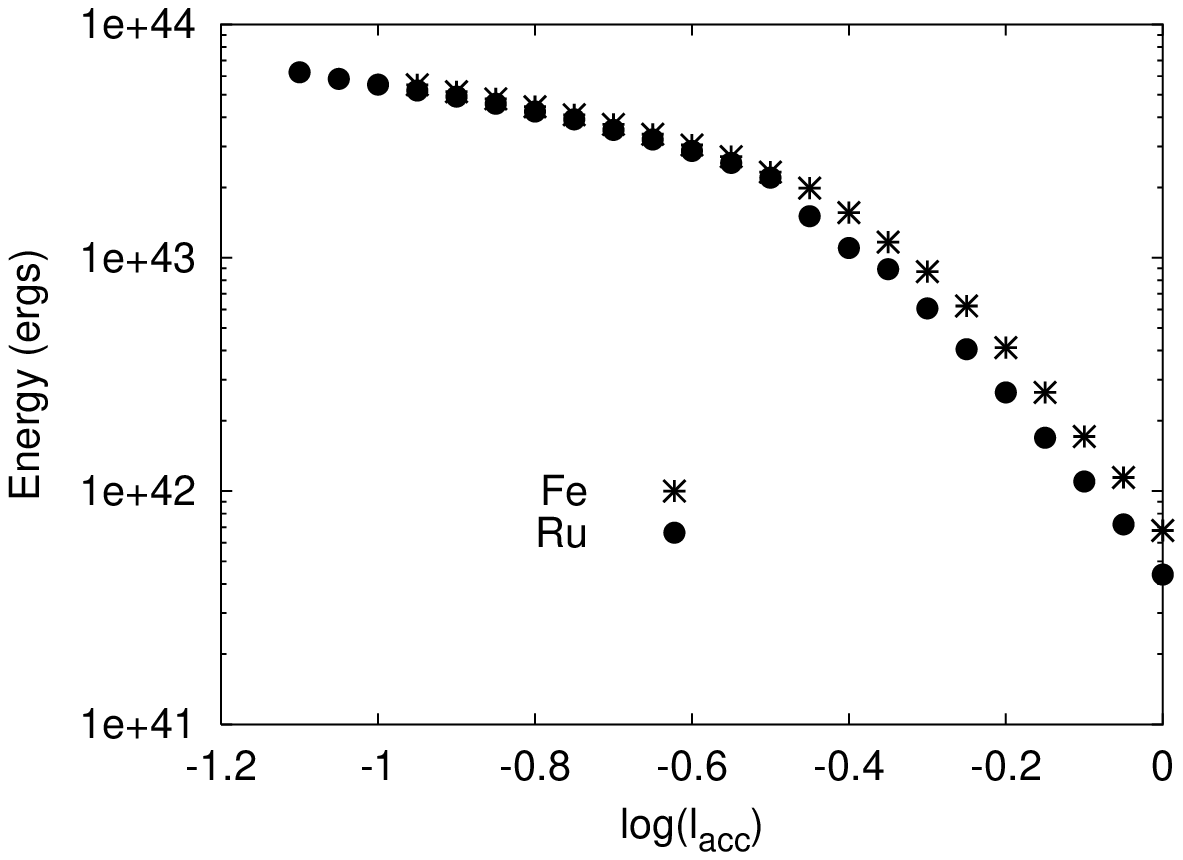}{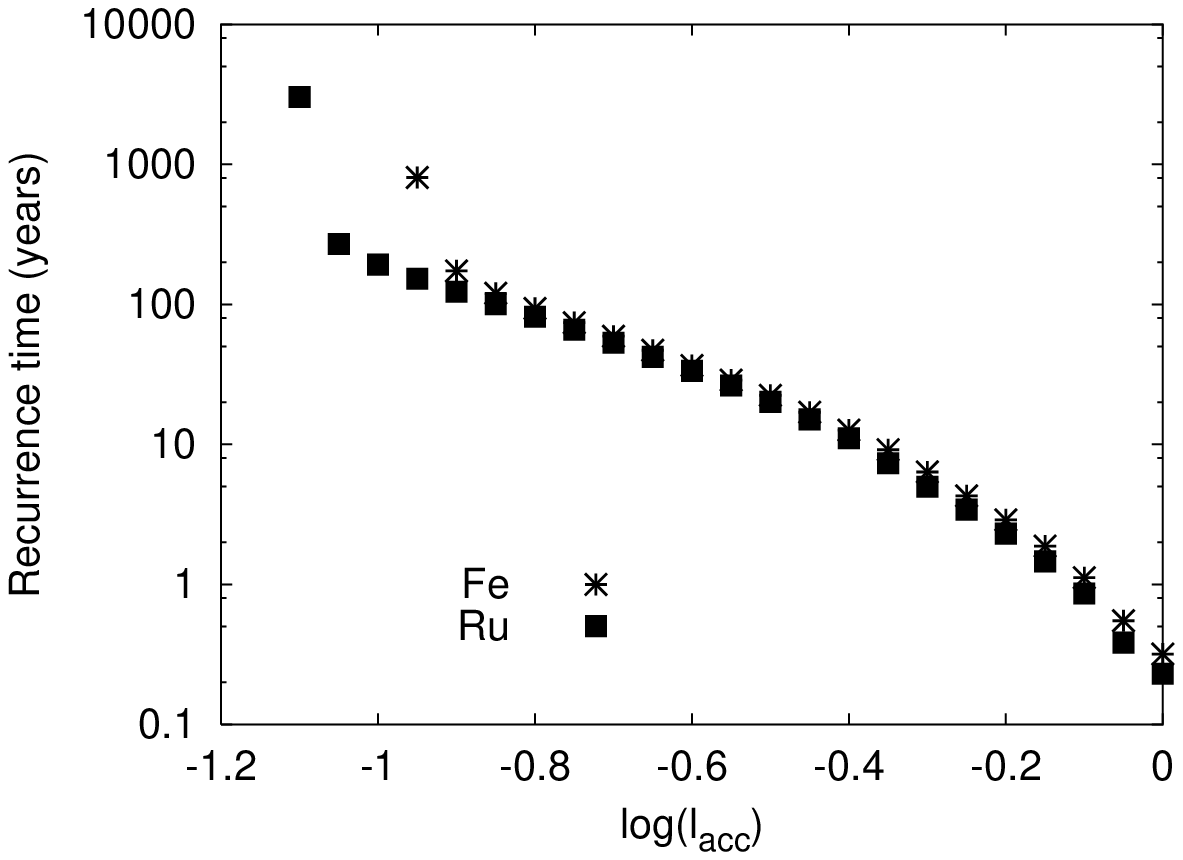}
\caption{Superburst energies and recurrence times for different heavy 
element compositions in the neutron star crust as a function of accretion 
rate.  The core is 
assumed to emit neutrinos via modified URCA reactions.
}
\label{fig10}
\end{figure}

\begin{figure}
\plottwo{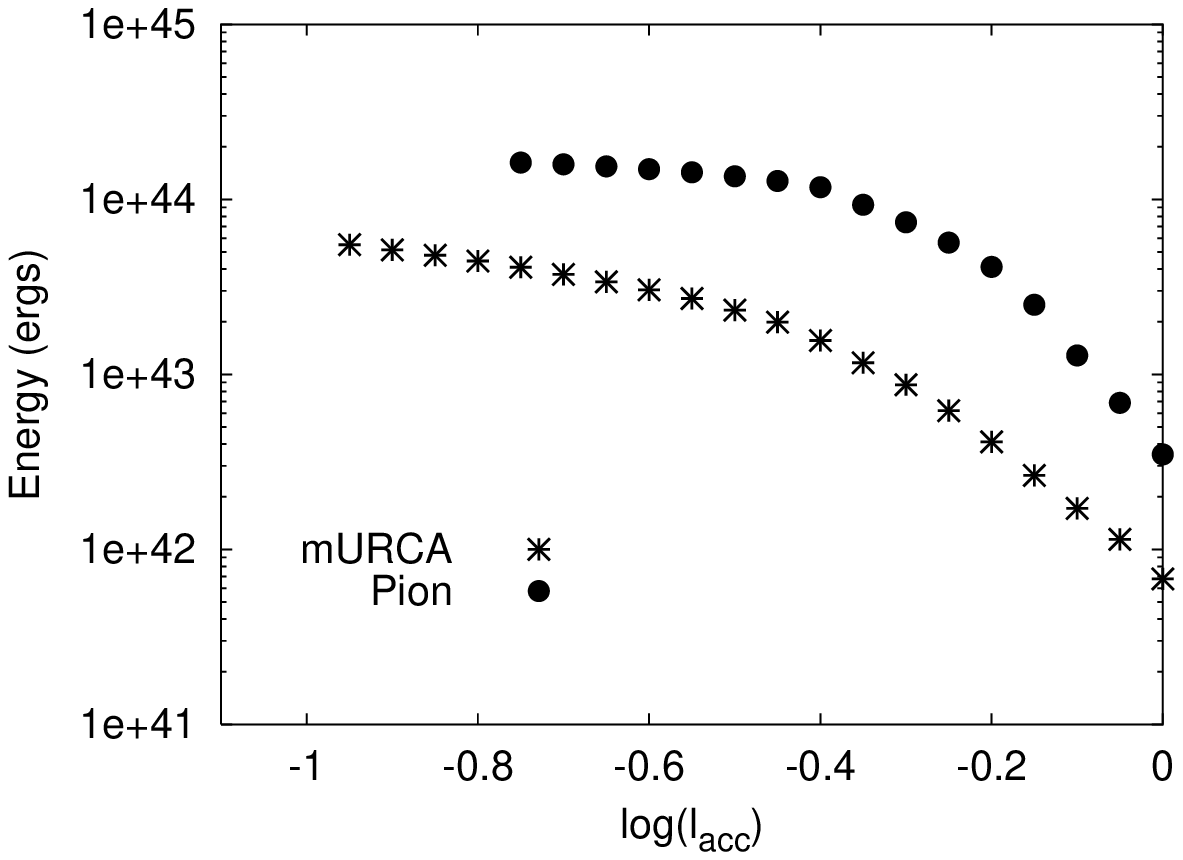}{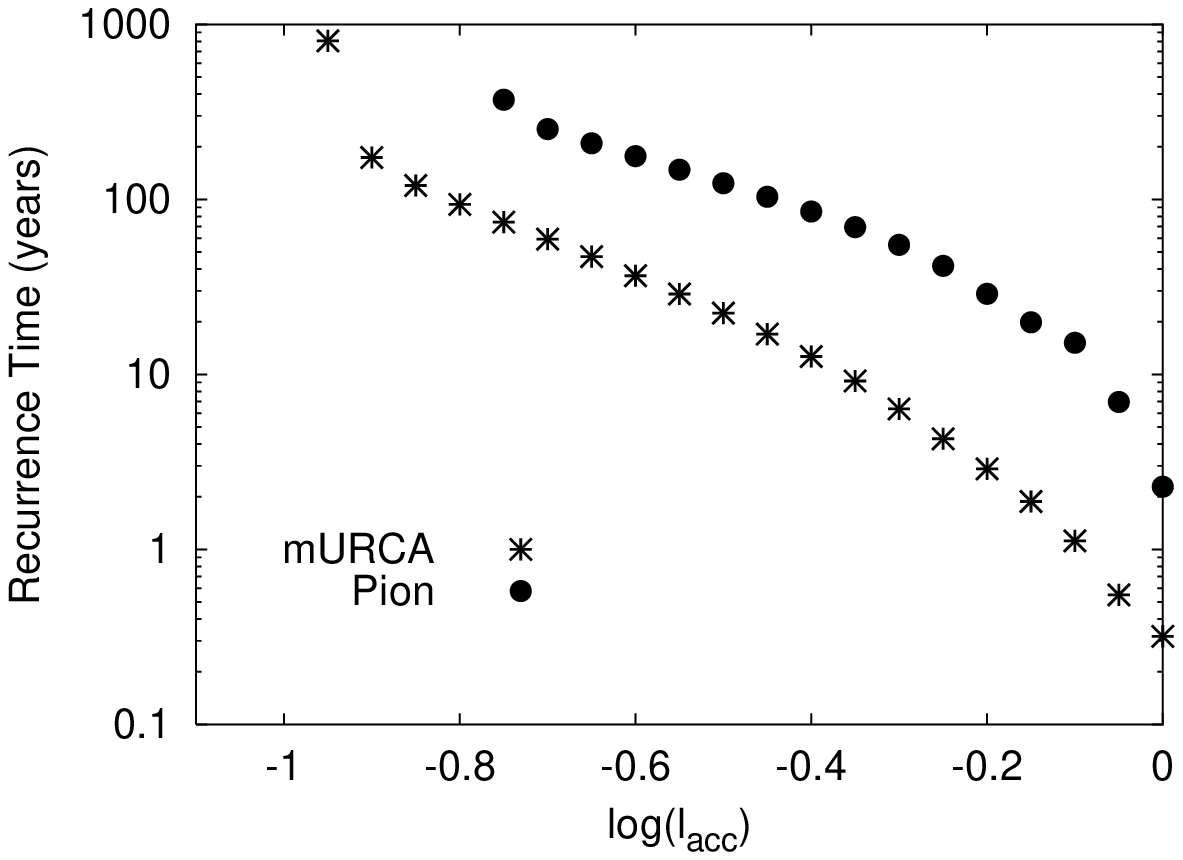}
\caption{Superburst energies and recurrence times for two core neutrino 
cooling mechanisms as 
a function of accretion rate.  The modified URCA and pion cooling models 
both assume a nonsuperfluid core.
}
\label{fig11}
\end{figure}

\begin{figure}
\plottwo{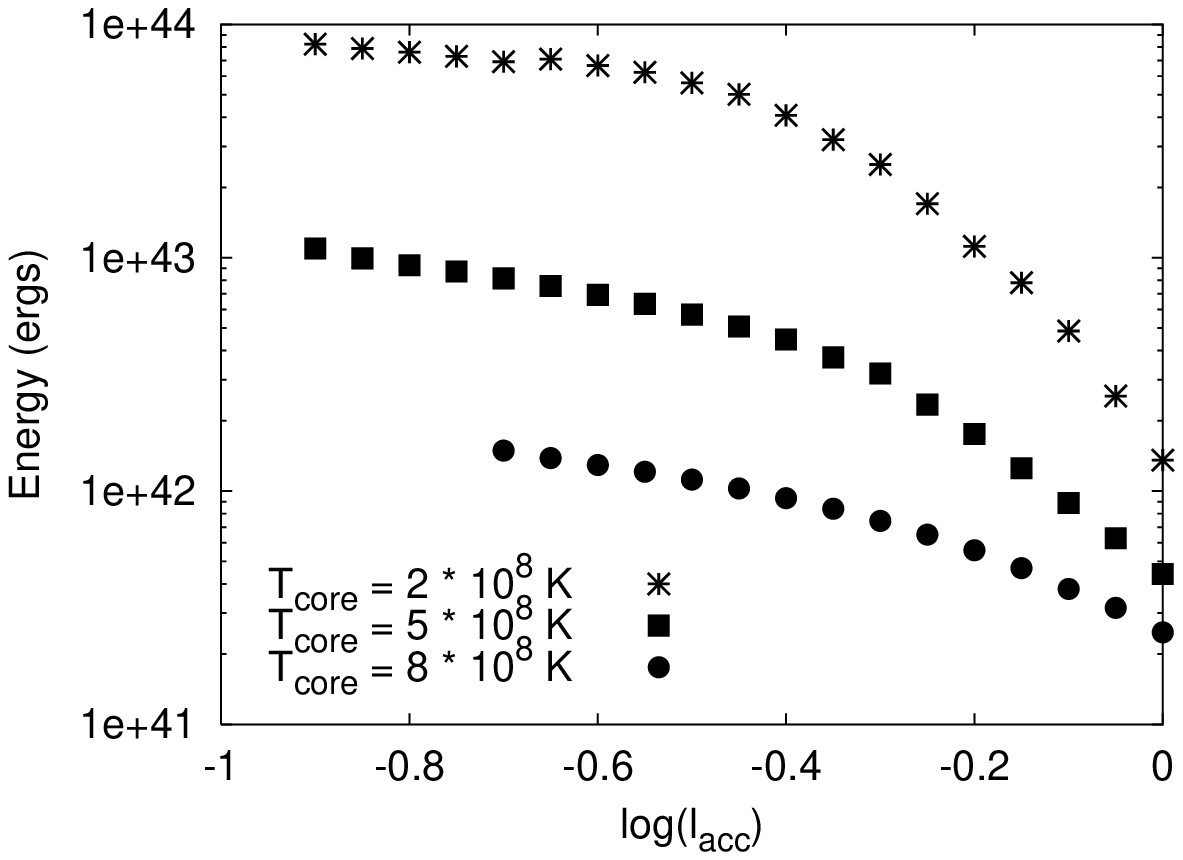}{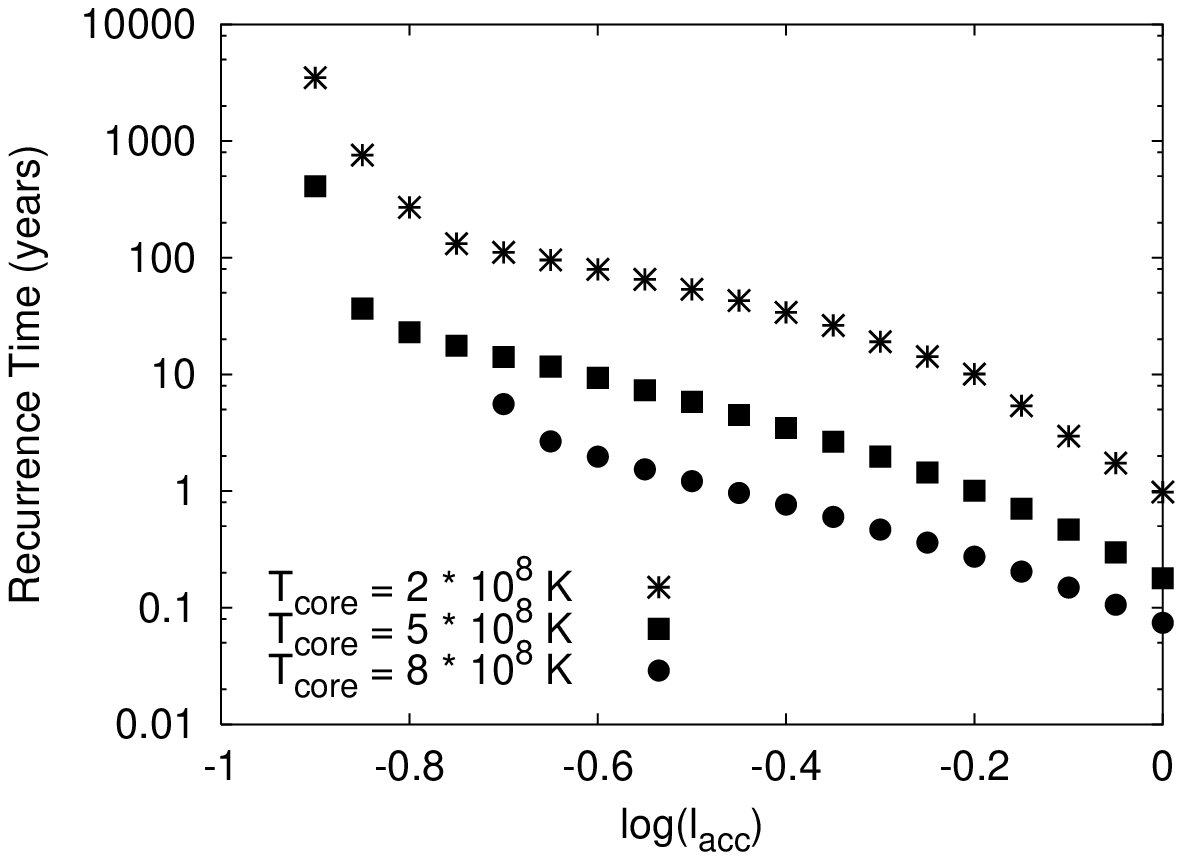}
\caption{Superburst energies and recurrence times for different fixed 
temperatures at the crust-core interface as a function of accretion rate.
}
\label{fig12}
\end{figure}

\begin{figure}
\plottwo{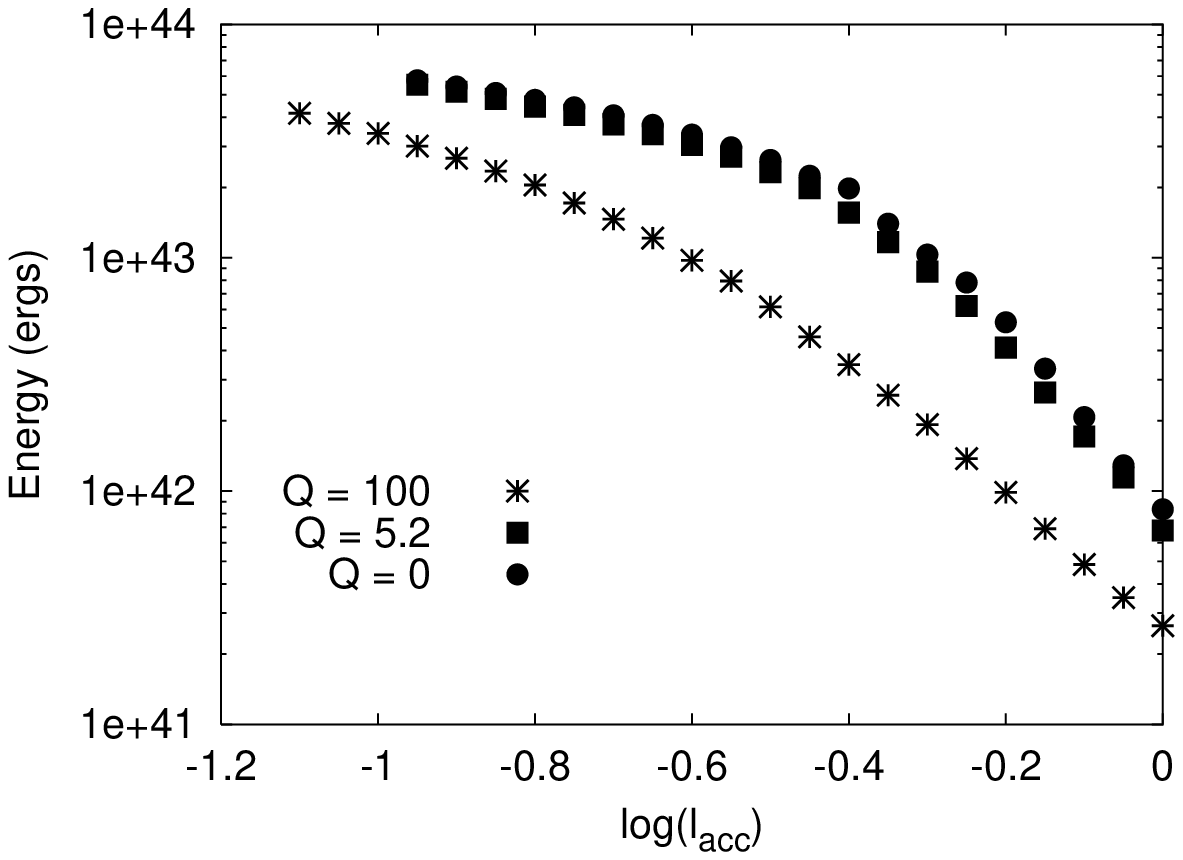}{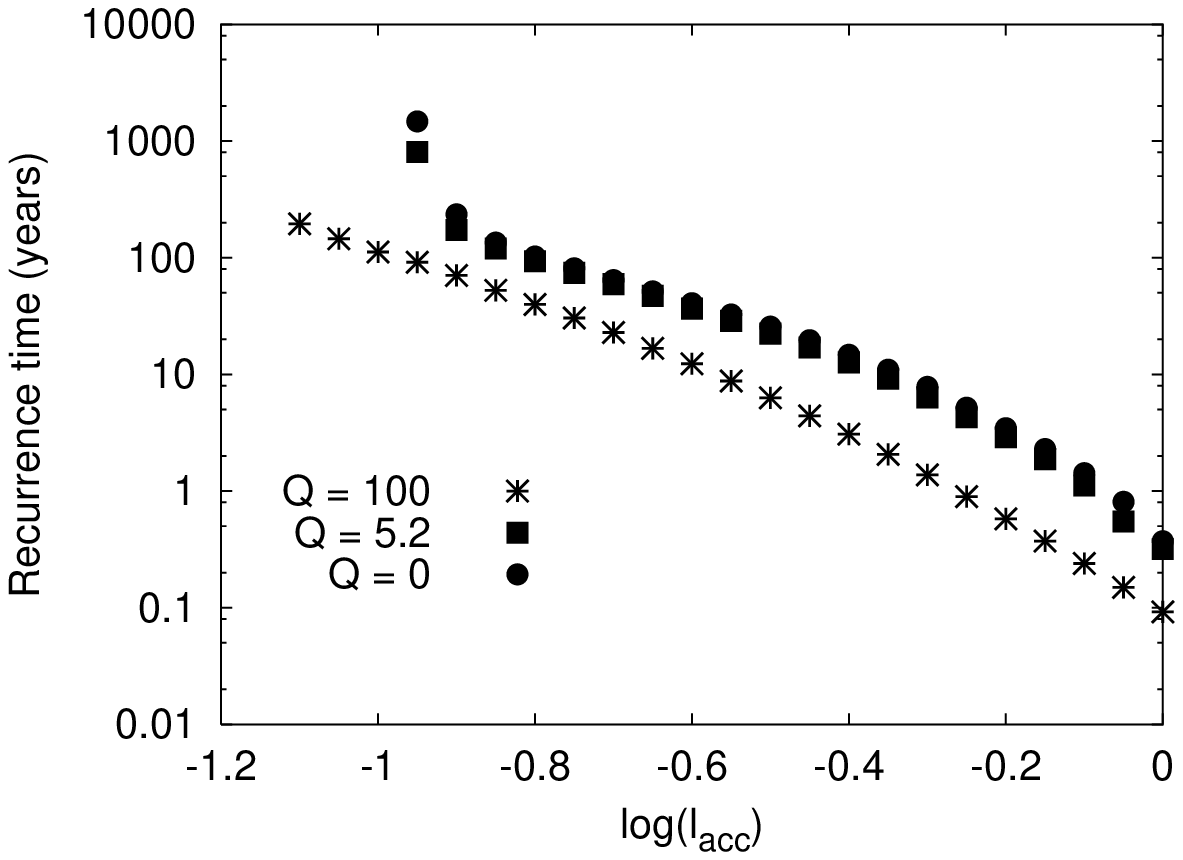}
\caption{Superburst energies and recurrence times for neutron stars with 
different impurity parameter values in the substrate, below the accreted 
layer.  The core is assumed to emit neutrinos via modified URCA reactions.
}
\label{fig13}
\end{figure}

\begin{figure}
\plottwo{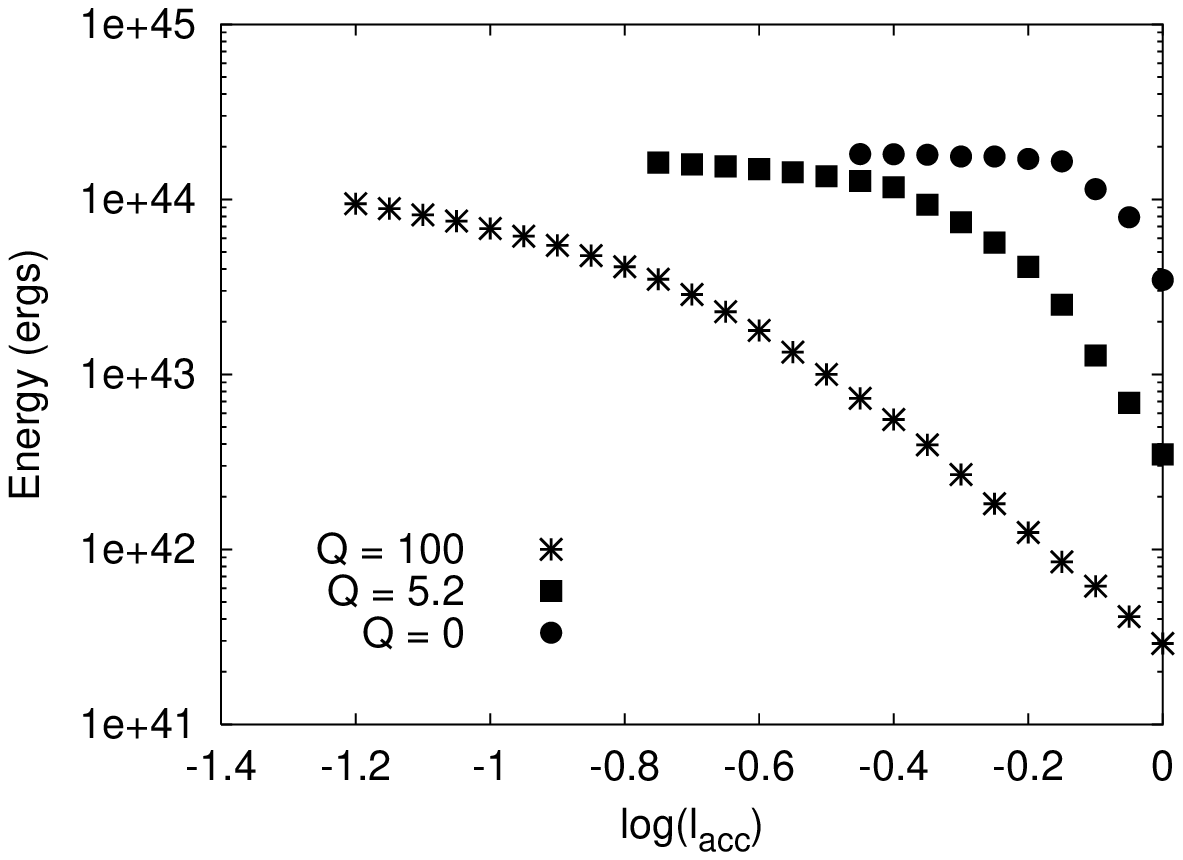}{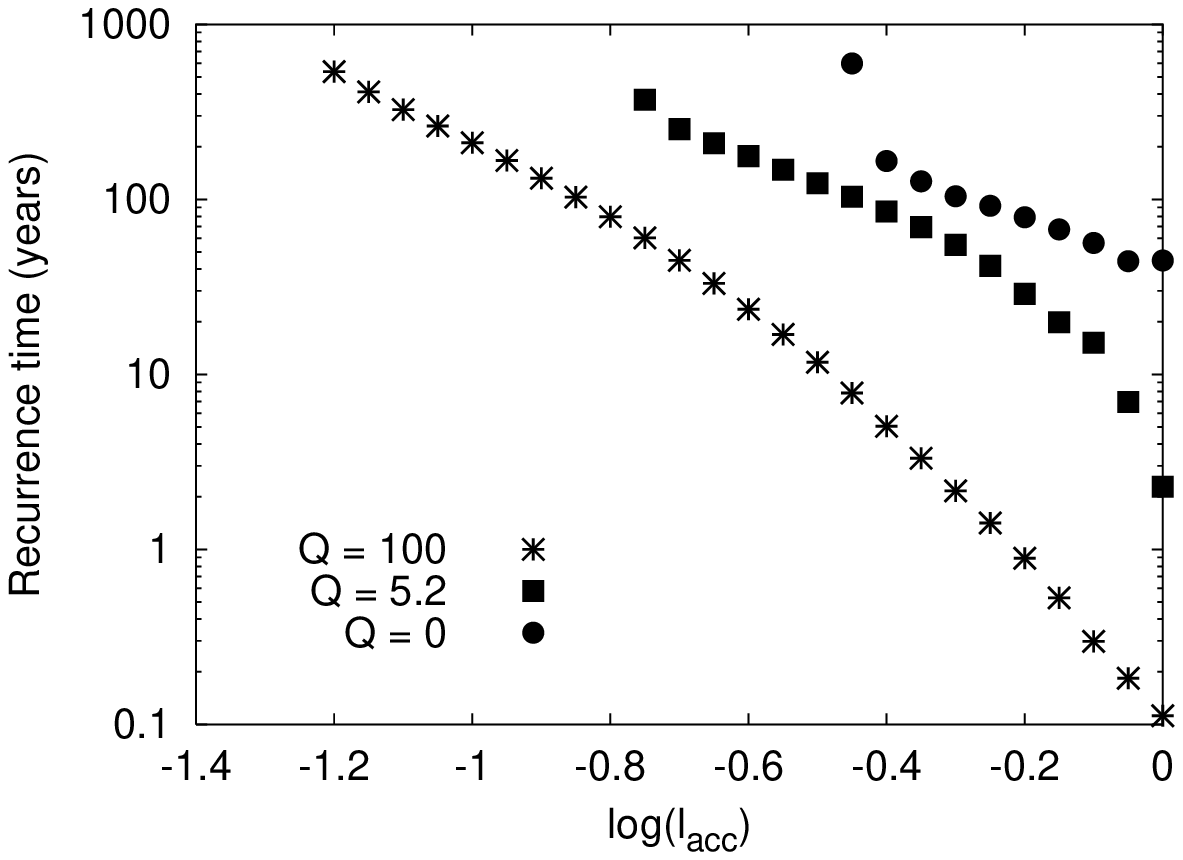}
\caption{Superburst energies and recurrence times for neutron stars with 
different impurity parameter values in the substrate, below the accreted 
layer.  The core is assumed to emit neutrinos via pionic reactions.
}
\label{fig14}
\end{figure}

\begin{figure}
\plottwo{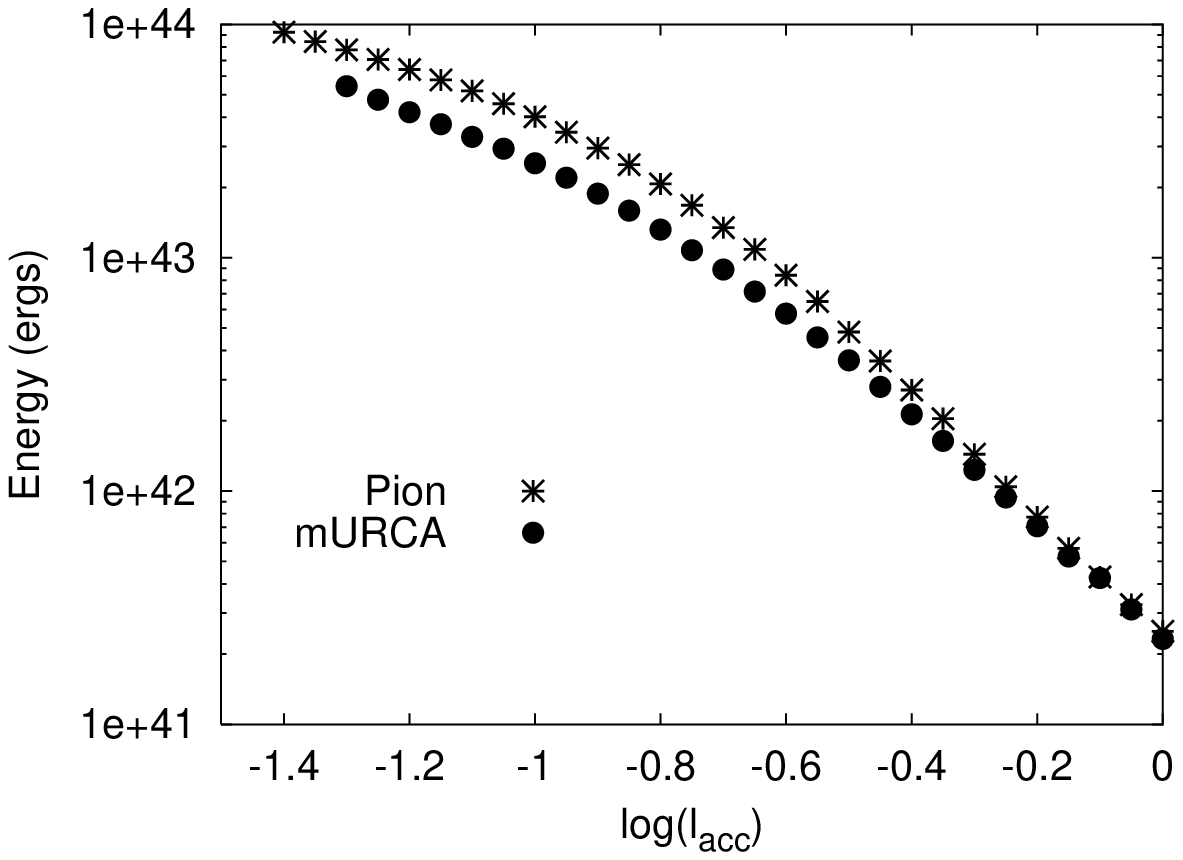}{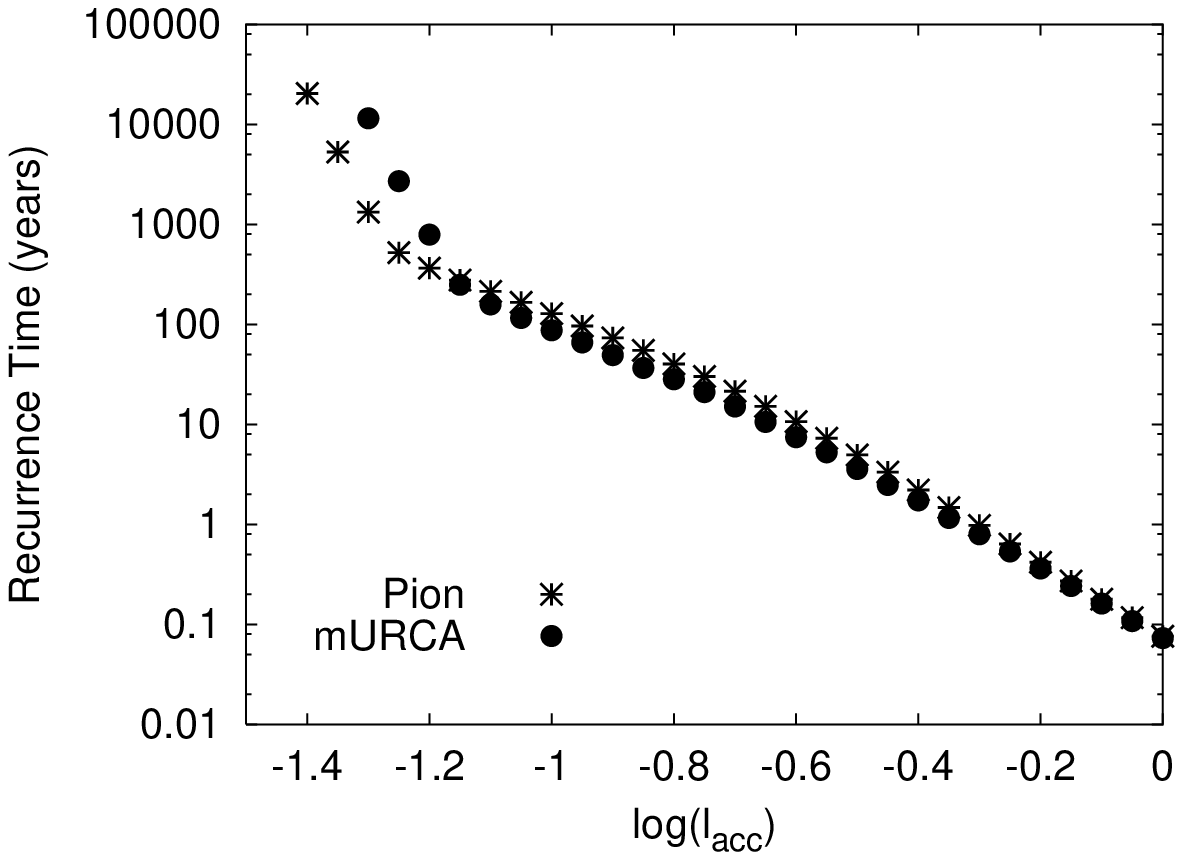}
\caption{Superburst energies and recurrence times for neutron stars with 
completely disordered crusts as a function of accretion rate.
}
\label{fig15}
\end{figure}

\clearpage

\begin{figure}
\plottwo{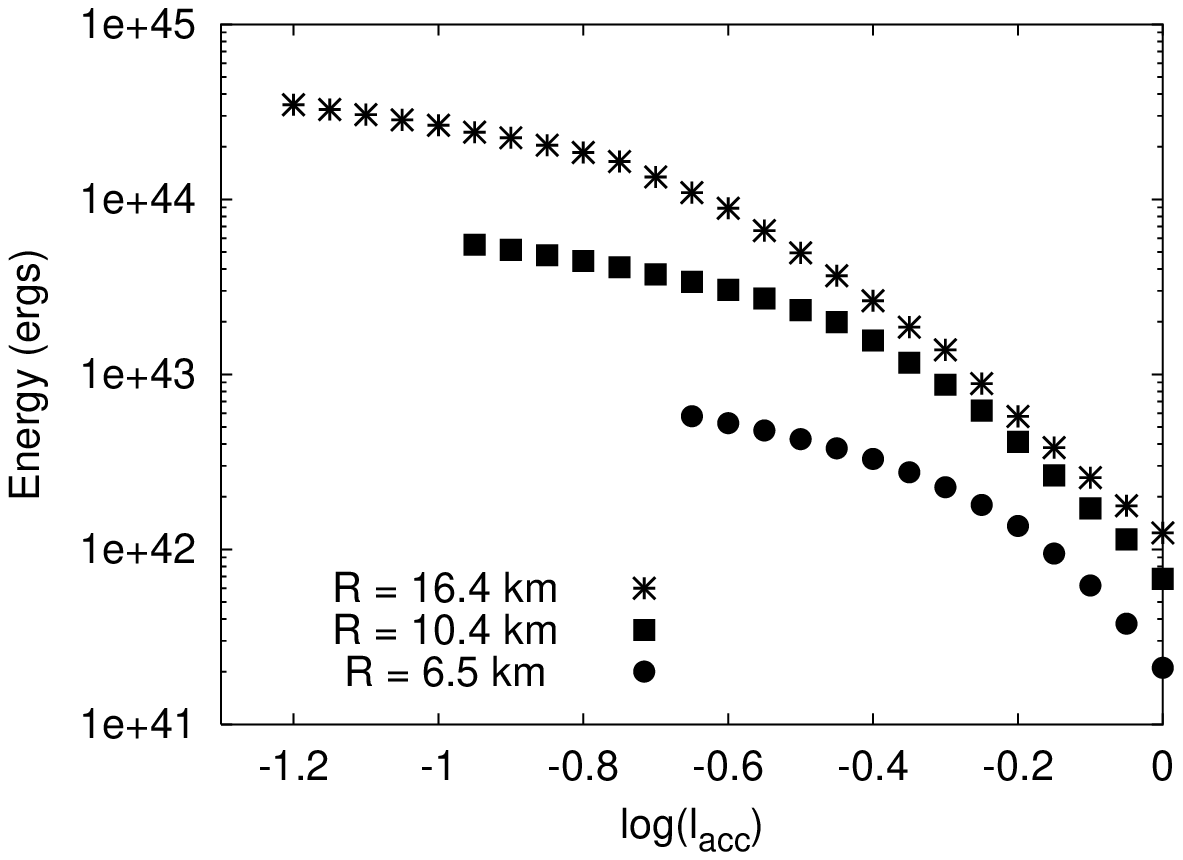}{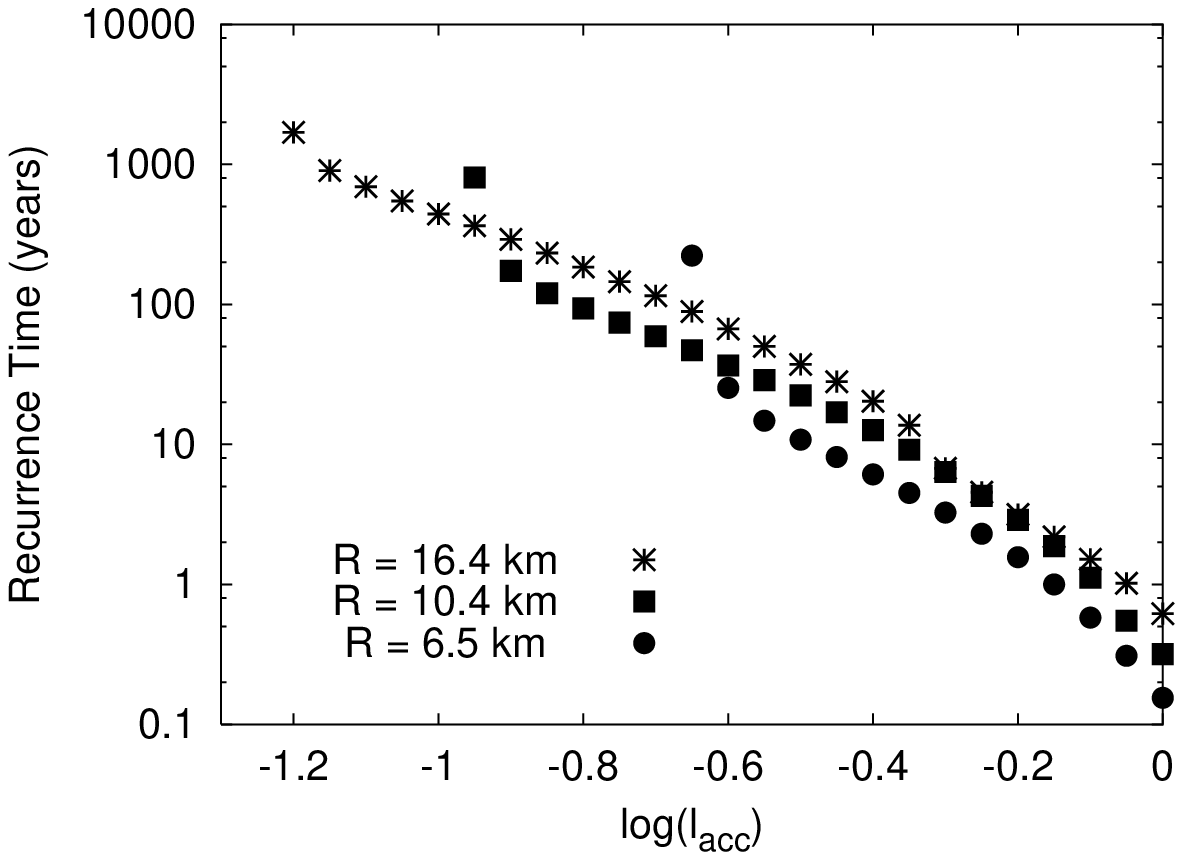}
\caption{Superburst energies and recurrence times for different stellar 
radii as a function of accretion rate.  The core is 
assumed to emit neutrinos via modified URCA reactions.
}
\label{fig16}
\end{figure}

\begin{figure}
\plottwo{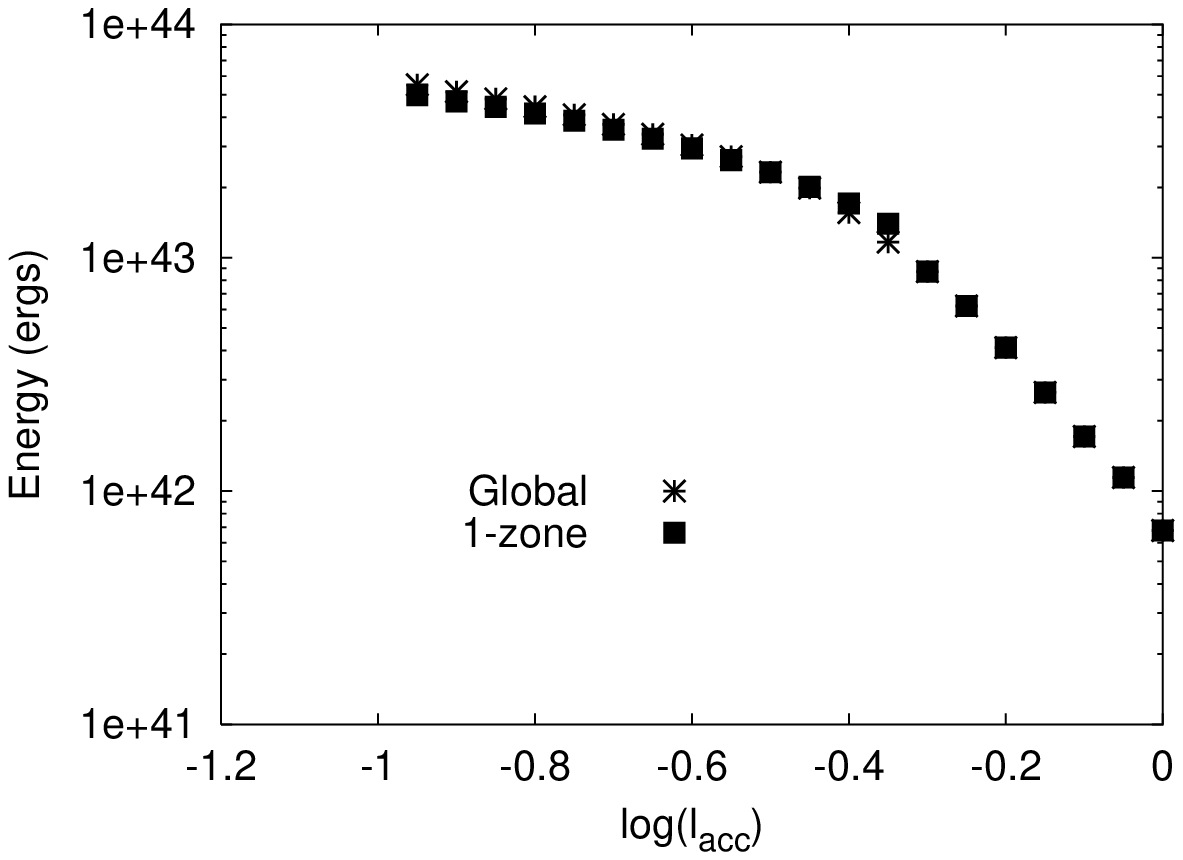}{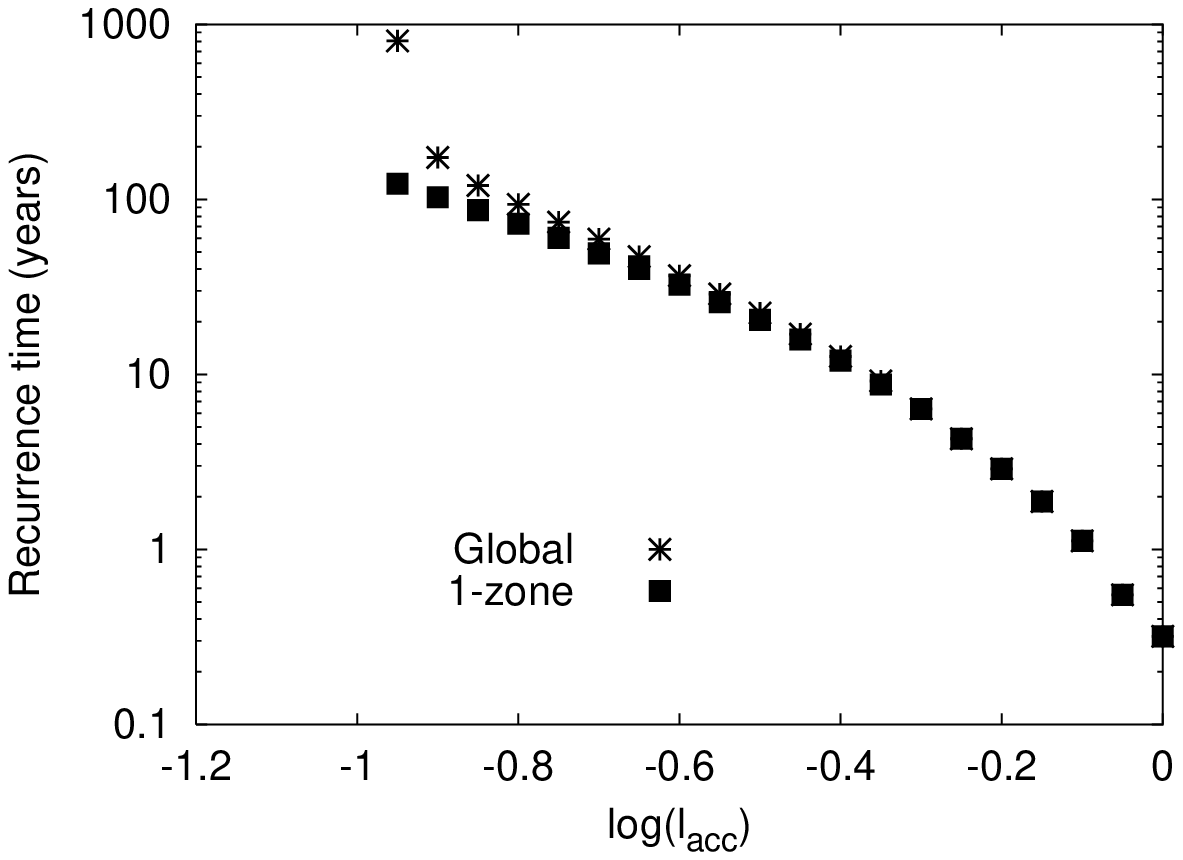}
\caption{A comparison of superburst energies and recurrence times between 
our global linear stability analysis and the local one-zone approximation.  
``Global'' signifies the results using our global linear stability analysis 
and ``1-zone'' signifies the results using the 
one-zone approximation for which we set 
$\mathrm{d} \ln \epsilon_{\mathrm{C}}/ \mathrm{d} \ln T = 26$.  The 
results for which we calculate $\mathrm{d} \ln \epsilon_{\mathrm{C}}/ 
\mathrm{d} \ln T$ self-consistently are nearly identical.  
The core is assumed to emit neutrinos via modified URCA reactions.
}
\label{fig17}
\end{figure}

\begin{figure}
\plottwo{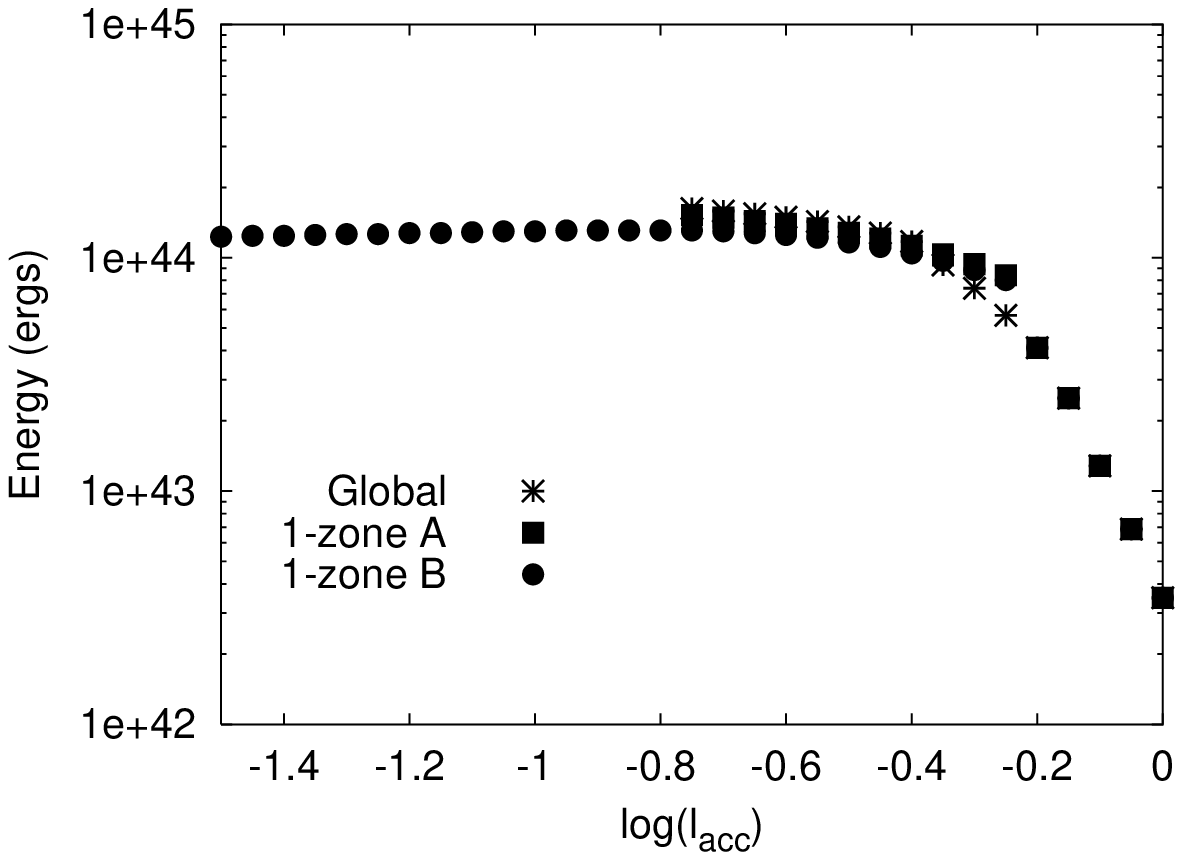}{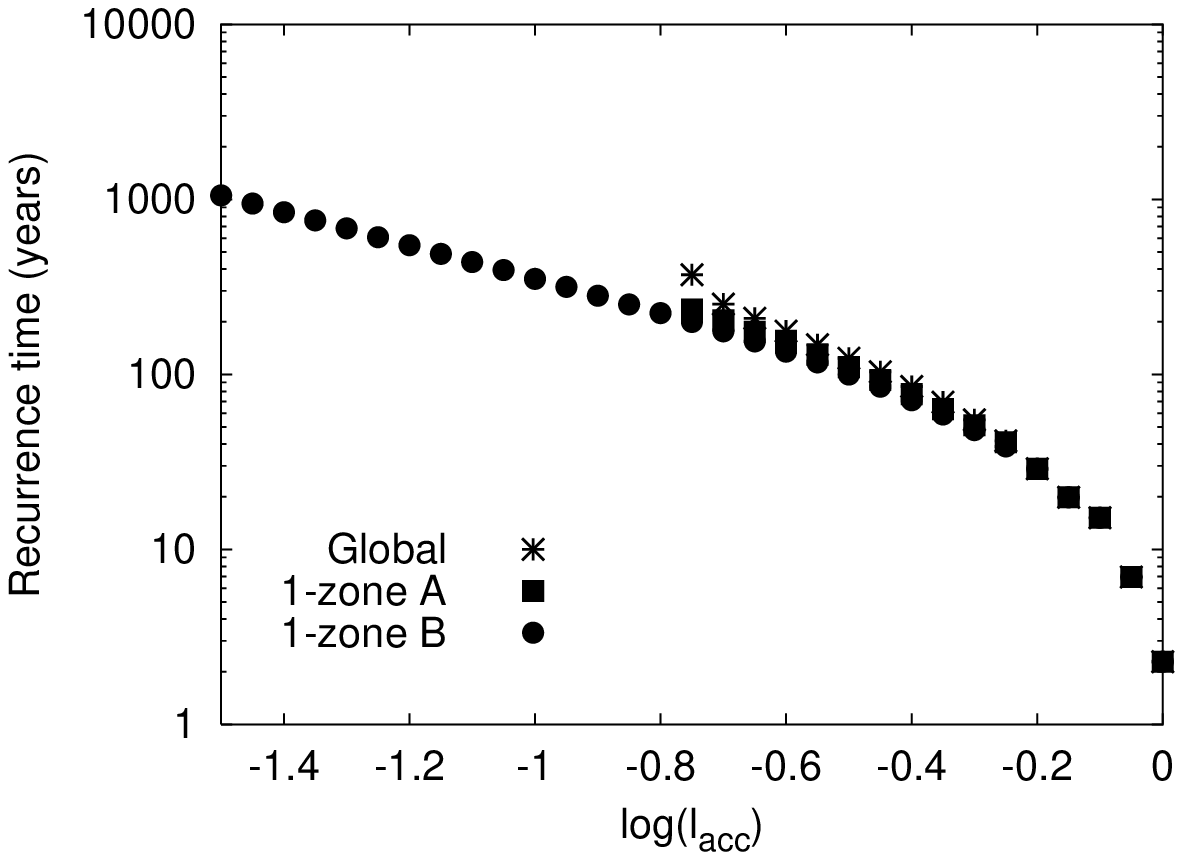}
\caption{A comparison of superburst energies and recurrence times between 
our global linear stability analysis and the local one-zone approximation.  
``Global'' signifies the results using our global linear stability analysis, 
``1-zone A'' signifies the results using the one-zone approximation for 
which we  calculate $\mathrm{d} \ln \epsilon_{\mathrm{C}}/ \mathrm{d} \ln T$ 
self-consistently (criterion i in \S6), and ``1-zone B'' signifies the 
results using the one-zone approximation for which we set 
$\mathrm{d} \ln \epsilon_{\mathrm{C}}/ \mathrm{d} \ln T = 26$ 
(criterion iii).
The core is assumed to emit neutrinos via pionic reactions.
}
\label{fig18}
\end{figure}

\end{document}